\documentclass[twocolumn]{aastex631}

\shorttitle{CEERS Mergers}
\shortauthors{Rose et al.}

\begin{document}

\title{CEERS Key Paper. IX. Identifying Galaxy Mergers in CEERS NIRCam Images Using Random Forests and Convolutional Neural Networks}

\author[0000-0002-8018-3219]{Caitlin Rose}
\author[0000-0001-9187-3605]{Jeyhan S. Kartaltepe}
\affil{Laboratory for Multiwavelength Astrophysics, School of Physics and Astronomy, Rochester Institute of Technology, 84 Lomb Memorial Drive, Rochester, NY 14623, USA}

\author{Gregory F. Snyder}
\affil{Space Telescope Science Institute,
3700 San Martin Dr., Baltimore, MD 21218, USA}

\author[0000-0002-1416-8483]{Marc Huertas-Company}
\affil{Instituto de Astrof\'isica de Canarias, La Laguna, Tenerife, Spain}
\affil{Universidad de la Laguna, La Laguna, Tenerife, Spain}
\affil{Universit\'e Paris-Cit\'e, LERMA - Observatoire de Paris, PSL, Paris, France}

\author[0000-0003-3466-035X]{{L. Y. Aaron} {Yung}}
\affil{Space Telescope Science Institute, 3700 San Martin Dr., Baltimore, MD 21218, USA}

\author[0000-0002-7959-8783]{Pablo Arrabal Haro}
\affil{NSF's National Optical-Infrared Astronomy Research Laboratory, 950 N. Cherry Ave., Tucson, AZ 85719, USA}

\author[0000-0002-9921-9218]{Micaela B. Bagley}
\affil{Department of Astronomy, The University of Texas at Austin, Austin, TX, USA}

\author[0000-0003-0492-4924]{Laura Bisigello}
\affiliation{INAF, Istituto di Radioastronomia, Via Piero Gobetti 101, 40129 Bologna, Italy}
\affiliation{Dipartimento di Fisica e Astronomia "G.Galilei", Universit\'a di Padova, Via Marzolo 8, I-35131 Padova, Italy}

\author[0000-0003-2536-1614]{Antonello Calabr{\`o}} 
\affil{INAF -- Osservatorio Astronomico di Roma, via di Frascati 33, 00078 Monte Porzio Catone, Italy}

\author[0000-0001-7151-009X]{Nikko J. Cleri}
\affil{Department of Physics and Astronomy, Texas A\&M University, College Station, TX, 77843-4242 USA}
\affil{George P.\ and Cynthia Woods Mitchell Institute for Fundamental Physics and Astronomy, Texas A\&M University, College Station, TX, 77843-4242 USA}

\author[0000-0001-5414-5131]{Mark Dickinson}
\affil{NSF's National Optical-Infrared Astronomy Research Laboratory, 950 N. Cherry Ave., Tucson, AZ 85719, USA}

\author[0000-0001-7113-2738]{Henry C. Ferguson}
\affil{Space Telescope Science Institute, 3700 San Martin Dr., Baltimore, MD 21218, USA}

\author[0000-0001-8519-1130]{Steven L. Finkelstein}
\affil{Department of Astronomy, The University of Texas at Austin, Austin, TX, USA}

\author[0000-0003-3820-2823]{Adriano Fontana}
\affil{INAF -- Osservatorio Astronomico di Roma, via di Frascati 33, 00078 Monte Porzio Catone, Italy}

\author[0000-0002-5688-0663]{Andrea Grazian}
\affil{INAF -- Osservatorio Astronomico di Padova, Vicolo dell'Osservatorio 5, I-35122, Padova, Italy}

\author[0000-0001-9440-8872]{Norman A. Grogin}
\affil{Space Telescope Science Institute, 3700 San Martin Dr., Baltimore, MD 21218, USA}

\author[0000-0002-4884-6756]{Benne W. Holwerda}
\affil{Physics \& Astronomy Department, University of Louisville, 40292 KY, Louisville, USA}

\author[0000-0001-9298-3523]{Kartheik G. Iyer}
\affil{Dunlap Institute for Astronomy \& Astrophysics, University of Toronto, Toronto, ON M5S 3H4, Canada}

\author[0000-0001-8152-3943]{Lisa J. Kewley}
\affil{Center for Astrophysics | Harvard \& Smithsonian, 60 Garden Street, Cambridge, MA 02138, USA}

\author[0000-0002-5537-8110]{Allison Kirkpatrick}
\affil{Department of Physics and Astronomy, University of Kansas, Lawrence, KS 66045, USA}

\author[0000-0002-8360-3880]{Dale D. Kocevski}
\affil{Department of Physics and Astronomy, Colby College, Waterville, ME 04901, USA}

\author[0000-0002-6610-2048]{Anton M. Koekemoer}
\affil{Space Telescope Science Institute, 3700 San Martin Dr., Baltimore, MD 21218, USA}

\author[0000-0003-3130-5643]{Jennifer M. Lotz}
\affil{Gemini Observatory/NSF's National Optical-Infrared Astronomy Research Laboratory, 950 N. Cherry Ave., Tucson, AZ 85719, USA}

\author[0000-0003-1581-7825]{Ray A. Lucas}
\affil{Space Telescope Science Institute, 3700 San Martin Dr., Baltimore, MD 21218, USA}

\author[0000-0002-8951-4408]{Lorenzo Napolitano}
\affil{INAF -- Osservatorio Astronomico di Roma, via di Frascati 33, 00078 Monte Porzio Catone, Italy}
\affil{Dipartimento di Fisica, Università di Roma Sapienza, Città Universitaria di Roma - Sapienza, Piazzale Aldo Moro, 2, 00185, Roma, Italy}

\author[0000-0001-7503-8482]{Casey Papovich}
\affil{Department of Physics and Astronomy, Texas A\&M University, College Station, TX, 77843-4242 USA}
\affil{George P.\ and Cynthia Woods Mitchell Institute for Fundamental Physics and Astronomy, Texas A\&M University, College Station, TX, 77843-4242 USA}

\author[0000-0001-8940-6768]{Laura Pentericci}
\affil{INAF -- Osservatorio Astronomico di Roma, via di Frascati 33, 00078 Monte Porzio Catone, Italy}

\author[0000-0003-4528-5639]{Pablo G. P\'erez-Gonz\'alez}
\affil{Centro de Astrobiolog\'{\i}a (CAB), CSIC-INTA, Ctra. de Ajalvir km 4, Torrej\'on de Ardoz, E-28850, Madrid, Spain}

\author[0000-0003-3382-5941]{Nor Pirzkal}
\affil{ESA/AURA Space Telescope Science Institute}

\author[0000-0002-5269-6527]{Swara Ravindranath}
\affil{Astrophysics Science Division, NASA Goddard Space Flight Center, 8800 Greenbelt Road, Greenbelt, MD 20771, USA}
\affil{Center for Research and Exploration in Space Science and Technology II, Department of Physics, Catholic University of America, 620 Michigan Ave N.E., Washington DC 20064, USA}

\author[0000-0002-6748-6821]{Rachel S. Somerville}
\affil{Center for Computational Astrophysics, Flatiron Institute, 162 5th Avenue, New York, NY, 10010, USA}

\author[0000-0002-4772-7878]{Amber N. Straughn}
\affil{Astrophysics Science Division, NASA Goddard Space Flight Center, 8800 Greenbelt Road, Greenbelt, MD 20771, USA}

\author[0000-0002-1410-0470]{Jonathan R. Trump}
\affil{Department of Physics, 196 Auditorium Road, Unit 3046, University of Connecticut, Storrs, CT 06269, USA}

\author[0000-0003-3903-6935]{Stephen M.~Wilkins}
\affil{Astronomy Centre, University of Sussex, Falmer, Brighton BN1 9QH, UK}
\affiliation{Institute of Space Sciences and Astronomy, University of Malta, Msida MSD 2080, Malta}

\author[0000-0001-8835-7722]{Guang Yang}
\affil{Nanjing Institute of Astronomical Optics and Technology, Nanjing 210042, China}

\begin{abstract}

A crucial yet challenging task in galaxy evolution studies is the identification of distant merging galaxies, a task which suffers from a variety of issues ranging from telescope sensitivities and limitations to the inherently chaotic morphologies of young galaxies. In this paper, we use random forests and convolutional neural networks to identify high-redshift JWST CEERS galaxy mergers. We train these algorithms on simulated $3<z<5$ CEERS galaxies created from the IllustrisTNG subhalo morphologies and the Santa Cruz SAM lightcone. We apply our models to observed CEERS galaxies at $3<z<5$. We find that our models correctly classify $\sim60-70\%$ of simulated merging and non-merging galaxies; better performance on the merger class comes at the expense of misclassifying more non-mergers. We could achieve more accurate classifications, as well as test for the dependency on physical parameters such as gas fraction, mass ratio, and relative orbits, by curating larger training sets. When applied to real CEERS galaxies using visual classifications as ground truth, the random forests correctly classified $40-60\%$ of mergers and non-mergers at $3<z<4$, but tended to classify most objects as non-mergers at $4<z<5$ (misclassifying $\sim70\%$ of visually-classified mergers). On the other hand, the CNNs tended to classify most objects as mergers across all redshifts (misclassifying $80-90\%$ of visually-classified non-mergers). We investigate what features the models find most useful, as well as characteristics of false positives and false negatives, and also calculate merger rates derived from the identifications made by the models.

\end{abstract}

\section{Introduction} \label{sec:ch3intro}

Galaxy merger studies are important for understanding the evolution of galaxies over cosmic time, as mergers can affect multiple physical phenomena and processes such as star formation and AGN activity \citep[e.g.,][]{ell2008,pat2011,lar2016}, the morphological evolution of spiral galaxies to elliptical galaxies \citep[e.g.,][]{too1977,cox2006,kor2009,rod2017}, and the build up of massive structures \citep[e.g.,][]{tacchella2015,zolotov2015, Costantin2021, Costantin2022}. To better understand galaxy evolution, it is necessary to be able to accurately identify merging galaxies over a wide range of mass ratios \citep{2010MNRAS.404..575L,2010MNRAS.404..590L} and redshifts.

However, at high redshift, merger identification can be difficult. The limited sensitivity of modern telescopes can result in samples biased toward the most massive galaxies and major mergers \citep[mass ratio $<$4:1; e.g.,][]{Ellison2013,man2018}. Cosmological surface brightness dimming and poor angular resolution can result in the loss of faint, detailed structures. As rest-frame optical light is shifted towards the infrared, infrared telescopes like JWST are needed to observe the stellar population. Finally, high redshift galaxies can inherently have irregular morphologies that are difficult to discern from merger signatures \citep[e.g.,][]{Dekel2009,Kartaltepe2012,kart2015}.

Classic methods of identifying mergers -- visual classifications \citep[e.g.,][]{lin2008, kart2015}, the close pair method \citep[e.g.,][]{lin2004,kart2007,ven2017,shah2020}, and quantitative morphology parameters \citep[e.g.,][]{con2003, lotz2004, free2013} -- all have limitations. The close pair method requires that both companions have accurate redshifts in order to identify pairs that are physically close in space, and is most sensitive to galaxies that have not yet merged. Visual classifications are qualitative and subjective, as well as time consuming. Finally, individual morphology parameters are each sensitive to different merger phases as well as noise \citep[e.g.,][]{2010MNRAS.404..575L,2010MNRAS.404..590L}. Recently, studies have begun investigating machine learning methods for merger identification tasks, as they show promise for exploiting complex data (whether from multi-band images or from pre-computed morphology parameters) to better detect merger signatures in high redshift images than classic methods \citep[e.g.,][]{sny2019,Ferreira2022}. Furthermore, the use of deep, high-resolution near-infrared JWST imaging will also help in the detection of merger features at high redshifts. One such JWST survey is the Cosmic Evolution Early Release Science Survey \citep[CEERS; ][]{fink2022}, which surveyed galaxies at $0.5 < z < 13$ over $100$ arcmin$^2$ in the Extended Groth Strip HST legacy field \citep[EGS; ][]{davis2007}. 

In \cite{rose2023}, we trained random forests to identify merging galaxies at $0.5 < z < 4$ in simulated CEERS images from IllustrisTNG, using a suite of morphology parameters measured using \texttt{Galapagos-2} \citep{bam2011,hau2013,vika2013} and \texttt{statmorph} \citep{rod2019}, as well as merger history information from IllustrisTNG. We found that the forests correctly classified $\sim$60\% of mock CEERS mergers and non-mergers across all redshift bins. In this follow-up study, we investigate using convolutional neural networks trained directly on the mock images for the task of merger identification at high redshift, as well as test our trained models on real galaxies from CEERS (described in \S \ref{sec:CEERSdescription}).

In recent years, many studies have trained neural networks to classify galaxies by morphology in general \citep[e.g.,][]{die2015,hue2018,hue2019,barchi2020,cheng2020,walm2022,2023PASA...40....1H,2023arXiv230502478H} or to identify mergers specifically \citep[e.g.,][]{ack2018,bot2019,pear2019,pearson2019_gm20,cip2020,cip2021,bick2021,fer2020,Ferreira2022}. However, most studies regarding merger identification train and test on galaxy samples at low redshift. At high redshift, \cite{cip2020,cip2021}, \cite{fer2020,Ferreira2022}, and \cite{pearson2019_gm20} each demonstrate that convolutional neural networks show promise for accurately identifying high redshift mergers. In particular, \cite{cip2020} apply their convolutional neural network (``DeepMerge") to $z=2$ simulated Illustris-1 HST F814W and F160W images created in \cite{sny2019} and achieve an accuracy of $76\%$ on the merger class for noisy data.

In this paper, we train DeepMerge\footnote{\href{https://github.com/deepskies/deepmerge-public}{https://github.com/deepskies/deepmerge-public}} to identify galaxy mergers from simulated CEERS images from IllustrisTNG. Then we explore using our trained random forest from \cite{rose2023} and DeepMerge algorithms to identify merging galaxies in real CEERS imaging. In \S \ref{sec:data}, we describe both the simulated JWST CEERS images and observed images, and how our datasets were curated. In \S \ref{sec:methods}, we describe training DeepMerge, as well as measuring quantitative parameters for observed galaxies. In \S \ref{sec:analysis} and \S \ref{sec:discuss}, we present our DeepMerge results on simulated data, then show our attempts to identify observed mergers using both DeepMerge and random forests. We summarize and conclude in \S \ref{sec:conclude}.

\section{Data} \label{sec:data}

\subsection{CEERS Observations and Data Reduction}\label{sec:CEERSdescription}

The Cosmic Evolution Early Release Science Survey \citep[CEERS; ][]{fink2022} is a JWST Early Release Science program (Proposal ID \#1345) that observed the EGS deep field in 10 NIRCam pointings. The first four NIRCam pointings -- hereafter CEERS1, CEERS2, CEERS3, and CEERS6 -- were taken on June 21, 2022 in seven filters (F115W, F150W, F200W, F277W, F356W, F410M, and F444W) with a typical exposure time of 2835 seconds per exposure. We use these four pointings in this work. A full description of CEERS and the data reduction steps can be found in \cite{fink2022} and \cite{bagley2023}, respectively. To summarize, the CEERS NIRCam images were reduced with version 1.7.2 of the JWST Calibration Pipeline\footnote{\href{jwst-pipeline.readthedocs.io}{jwst-pipeline.readthedocs.io}} \citep{bushouse2022} with custom modifications and with current NIRCam reference files\footnote{\href{jwst-crds.stsci.edu}{jwst-crds.stsci.edu}, jwst 0989.pmap, jwst nircam 0232.imap}. The images were processed through Stages 1 and 2 of the pipeline, where reduction steps included detector-level corrections; removal of $1/f$ noise, wisps, snowballs; and astrometric calibration. The images were distortion-corrected and combined into final mosaics in Stage 3 of the JWST pipeline using the drizzle algorithm with inverse variance map weighting \citep{fru2002,caser2000}. The final mosaics for each pointing in each filter are background-subtracted and have pixel scales of 0\farcs03/pixel. The publicly released images are available at \href{https://ceers.github.io/releases.html}{https://ceers.github.io/releases.html} and on MAST via doi: \dataset[10.17909/z7p0-8481]{https://dx.doi.org/10.17909/z7p0-8481}.

In this work, we use the sample of 850 galaxies at $z>3$, from the four CEERS pointings, that were selected by \cite{kart2023} as well as their visual classifications (described in \S \ref{sec:kartaltepe2023}).

\subsection{Simulated CEERS Images} \label{sec:simCEERS}

The simulated mock CEERS images used in this work were created by \cite{rose2023}. The images were constructed from IllustrisTNG100-1 \citep{TNG1Springel2018,TNG2Naiman2018,TNG3Nelson2018,TNG4Pillepich2018,TNG5Marinacci2018} and the Santa Cruz SAM lightcone \citep{som2021,2022MNRAS.515.5416Y}. We refer the reader to \cite{rose2023} for further details on the creation of the pristine, noiseless images. The images are available in six filters (F115W, F150W, F200W, F277W, F356W, F444W) and cover $\sim$ 100 square arcmin (with pixel scales of 0\farcs03/pixel) and contain over 100,000 galaxies, to mimic the size of the complete CEERS mosaic of the EGS field. The images were convolved with model PSFs, then noise was added on top to mimic CEERS noise \citep[estimated using the JWST Exposure Time Calculator;][]{pont2016}. The mock CEERS images are publicly available at \href{https://ceers.github.io/ancillary\_data.html}{https://ceers.github.io/ancillary\_data.html}.

\begin{figure}
\includegraphics[scale=0.26]{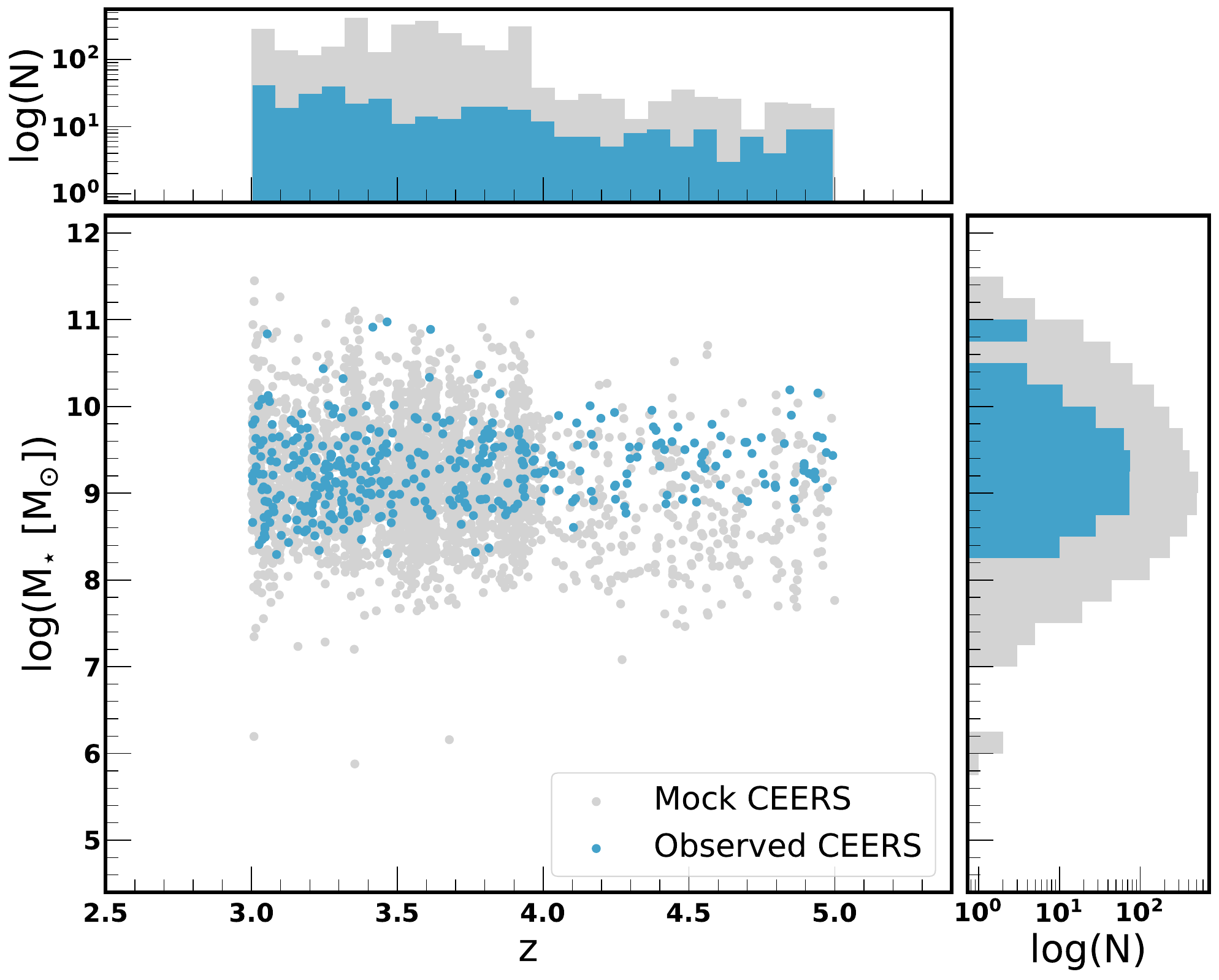}
\caption{Stellar mass versus redshift for objects from the mock CEERS sample at $3<z<5$ (\textit{grey}), and from the observed CEERS sample with $3<z<5$ (\textit{blue}). Above and to the right are the distributions of redshift and stellar mass, respectively. These samples do include the S/N and flag cuts discussed in \S \ref{sec:rf_training} and \S \ref{sec:visclass}.
\label{fig:mvz}}
\end{figure}

These images are accompanied by catalogs with simulation information such as redshift, star formation rate, and stellar mass. Additionally, the merger history catalogs for IllustrisTNG galaxies \citep{rod2015,nel2019} were also available. These catalogs give the IllustrisTNG snapshot numbers for each galaxy’s most recent past merger and next future merger (both major and minor). This merger history information makes IllustrisTNG an ideal dataset for training and testing machine learning algorithms to identify galaxy mergers at different stages. From the full simulated mosaic, we obtained morphology measurements for 5438 galaxies with $z > 3$. Figure \ref{fig:mvz} shows the redshift and mass distribution for these objects in gray. This sample also peaks at $z \sim 3$ and extends out to $z \sim 9$. This sample also contains more low-mass galaxies than in the observed CEERS sample. While \cite{TNG4Pillepich2018} note that the IllustrisTNG galaxy stellar mass function (GSMF) is in better agreement with observations than the Illustris GSMF, expecially at $z\leq1$, IllustrisTNG may still overproduce low-mass galaxies at higher redshifts. Stellar masses for the observed CEERS galaxies were computed by \cite{kart2023} using the EGS photometry \citep{stefanon2017} from CANDELS \citep{Gro2011,Koe2011} and estimated redshifts \citep{kodra2023}.

\section{Methods} \label{sec:methods}

\subsection{Visual Classifications of Observed Galaxies} \label{sec:kartaltepe2023}

Visual classifications were reported by \cite{kart2023}, and were carried out by members of the CEERS team. Images of the galaxies in our sample were hosted on the Zooniverse project builder\footnote{\href{https://www.zooniverse.org/lab}{ https://www.zooniverse.org/lab}}. Volunteer classifiers viewed stamps of the galaxies in seven JWST NIRCam filters and three HST  ACS/WFC3 filters, as well as RGB stamps and JWST segmentation map stamps \citep[see Figure 14 in the Appendix of][]{kart2023} from \texttt{Source Extractor} \citep[\texttt{SE} v2.23.2;][]{ber1996}. Each object was classified by three volunteer classifiers who followed a morphology classification scheme based on that used in \cite{kart2015}. The five tasks in this scheme ask volunteer classifiers to select main morphology classes, interaction classes, structural and quality flags, and, if the classifier wishes, to leave comments about the object. The main morphology class options were: Disk, Spheroid, Irregular / Peculiar, Point Source / Unresolved, and Unclassifiable / Junk. There were four interaction class options. 

\textit{(1) Merger}: Galaxies (that are single objects) that appear to have experienced a merger due to the presence of structures like tidal tails, double nuclei, or other asymmetries. All mergers should have Irregular/Peculiar selected as their main classification, but not all galaxies classified as irregulars are mergers.

\textit{(2) Interaction within \texttt{SE} segmentation map}: Two or more galaxies that appear to be interacting within the same segmentation map. There must be clear evidence of an interaction (tidal arms, bridges, etc.).

\textit{(3) Interaction beyond \texttt{SE} segmentation map}: Two or more galaxies that appear to be interacting that have their own distinct segmentation maps. 

\textit{(4) Non-interacting companion}: The main galaxy has a nearby companion, yet has no evidence of tidal interaction.

\begin{figure}
\includegraphics[scale=0.52]{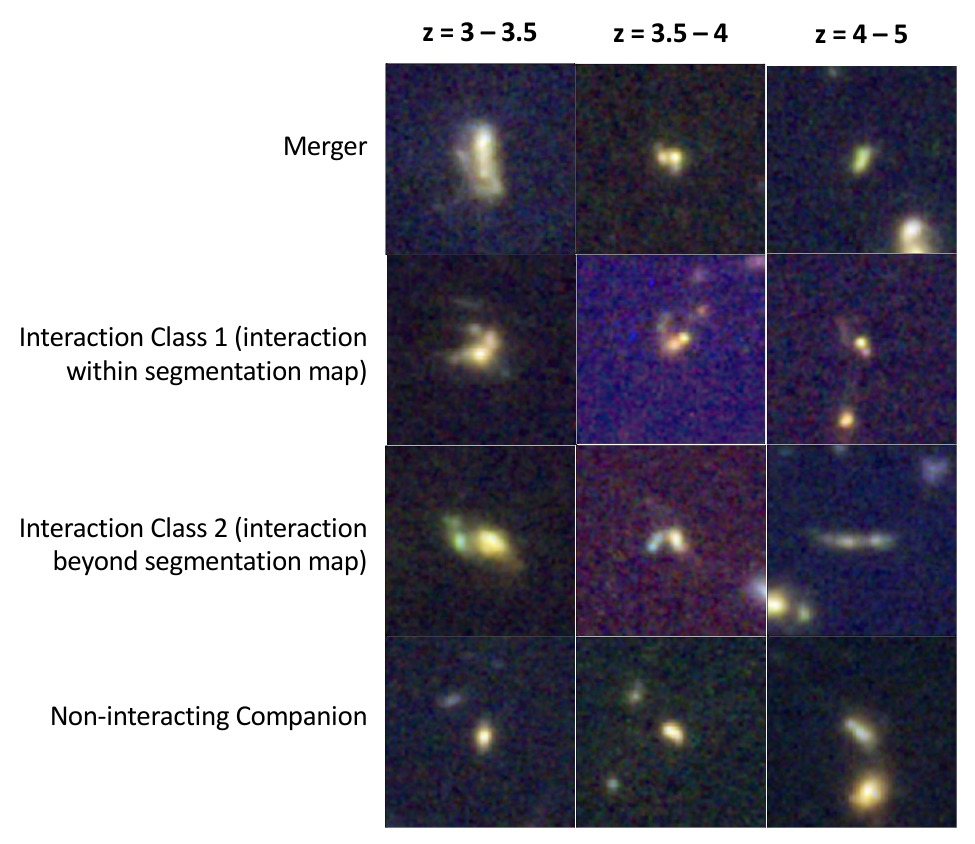}
\caption{Examples of CEERS galaxies (F150W+F200W+F356W stamps) for each visually-classified interaction class at $3.5 < z < 4$,  $3.5 < z < 4$, and $4 < z < 5$.
\label{fig:merger_examples}}
\end{figure}

The structural and quality flags question included flags to mark galaxies with merger features such as tidal tails and double nuclei. Figure \ref{fig:merger_examples} shows examples of galaxies from the different interaction classes.

\subsection{Morphology Parameters for Observed Galaxies} \label{sec:obs_morph_params}

\texttt{SE} \citep{ber1996} was used to detect $z>3$ galaxies in each of the four pointings. Weighted F150W$+$F200W$+$F277W$+$F356W images were used as the detection images. We ran \texttt{SE} in both ``cold" and ``hot" mode, in order to deblend overlapping galaxies and detect faint galaxies, respectively. The ``cold" and ``hot" detections were then combined following the logic of the \texttt{Galapagos-2} program \citep{bam2011,bar2012,hau2013,vika2013}.

\texttt{Galapagos-2} (Galaxy Analysis over Large Areas: Parameter Assessment by \texttt{Galfit}-ting Objects from \texttt{SE}) is an IDL program that runs the multiwavelength S\'ersic fitting program \texttt{GalfitM} \citep{peng2002,peng2010,peng2012} on large survey images. Information from the output \texttt{SE} catalog was used as initial guesses for the S\'ersic fits from \texttt{Galapagos-2}. Empirical point-spread functions (PSFs) made from stacking stars \citep{fink2022} were used for these fits. We ran \texttt{Galapagos-2} on all seven filters simultaneously to obtain S\'ersic indices and related size and magnitude information. In addition to the output S\'ersic model catalog, \texttt{Galapagos-2} also returns input image, model, and S\'ersic residual stamps for each galaxy in the catalog.

We also ran the non-parametric morphology program \texttt{statmorph} \citep{rod2019} separately on each filter for each of the four pointings, for the $z>3$ objects in the \texttt{Galapagos-2} catalog, in order to obtain morphology parameters such as concentration ($C$), asymmetry ($A$), and clumpiness ($S$) \citep{ber2000,con2000,con2003}, and the Gini coefficient ($G$) and the moment of light ($M_{20}$) \citep{abra2003,lotz2004}. The output \texttt{SE} segmentation map was used as input to \texttt{statmorph} to aid in calculating these parameters, as well as the PSFs from \cite{fink2022}.

Finally, it has been shown that asymmetric or unusual features indicative of mergers can be more prominent in S\'ersic residuals \citep[e.g.,][]{man2019}. Therefore, we also run \texttt{statmorph} on the S\'ersic residuals output by \texttt{Galapagos-2}. To do so, we add an offset of $+1$ to every pixel of all residual images, since \texttt{statmorph} requires that input images have positive flux which is not necessarily true for residual images.

We begin with the catalog of visual classifications for 850 objects at $z>3$ from \cite{kart2023}. After combining all of the above catalogs (matching by CANDELS ID number or by RA and Dec), we create a sample of 798 observed $z>3$ CEERS galaxies that have successful morphology measurements from both the \texttt{statmorph} and \texttt{Galapagos} catalogs. 

\subsection{Training Random Forests on Simulated Galaxies} \label{sec:rf_training}

In \cite{rose2023}, we trained and tested the random forest algorithm \citep{ho1995,brei2001} on the simulated images described in \S \ref{sec:simCEERS}. To do so, we ran \texttt{Galapagos-2} and \texttt{statmorph} on the mock CEERS data in a similar fashion as described in \S \ref{sec:obs_morph_params}. These morphology parameters were used as the input ``features" for the forests, which were the single S\'ersic index $n$, the two-component S\'ersic indices $n_{bulge}$ and $n_{disk}$, and various \texttt{statmorph} parameters \citep[outlined in ][]{rose2023}, in all six filters, for both science and residual images. From the merger history catalogs, we convert the IllustrisTNG snapshot numbers to ages of the Universe in Gyr, then calculate the difference in time between the galaxy of interest and the last and next merger events. The resulting time since the last merger (both major and minor) and the time until the next (both major and minor) are used as the ``labels" for the random forests. ``Mergers" are objects that had, or will have, a merging event within $\pm 250$ Myr, in accordance with \cite{sny2017}, as merger signatures from outside that time frame will have likely vanished. ``Non-mergers" are objects which either have a merging event outside that time frame, or have not had merging events at all. In \cite{rose2023}, we restrict the mock CEERS dataset to have \texttt{statmorph} signal-to-noise per pixel (S/N$_{F115W}$) $> 3$ and \texttt{Flag}$_{\texttt{Galfit}}$ $=2.0$ (which indicates successful Galfit measurements). Then, we split the final sample of 40,391 mock galaxies into seven redshift bins from $0.5 < z < 4$ and trained a separate forest for each bin, using \texttt{BalancedRandomForestClassifier()} from Python's \texttt{imblearn} package as the random forest algorithm and \texttt{keras-tuner} \citep{omal2019} for hyperparameter optimization.

Following the same procedure in \cite{rose2023}, we additionally train a random forest for $z = 4 - 5$. In \S \ref{sec:rf_sim}, we describe the outcome of training and testing random forests for merger identification on the mock CEERS dataset. In \S \ref{sec:rf_real}, we apply the $3 < z < 3.5$, $3.5 < z < 4$, and $4 < z < 5$ random forests to observed CEERS galaxies.

Additionally, we compute the Spearman correlations between the morphology parameters and the merger timescales using \texttt{spearmanr} from the \texttt{SciPy} package in order to understand which features might be most important for the random forests. The Spearman correlation describes the relationship between variables, where $+1$ or $-1$ indicates perfect positive or negative correlations, respectively, and 0 indicates no correlation. For each of the redshift bins studied in this work ($3 < z < 3.5$, $3.5 < z < 4$, and $4 < z < 5$), the strongest correlations ranged from $\sim |0.1| - |0.2|$. For example, the most correlated features for the $4 < z < 5$ bin were: F356W concentration with a correlation coefficient of $-0.21$, F200W residual flux (within an elliptical aperture) with a correlation coefficient of $0.11$, F444W asymmetry with a correlation coefficient of $0.27$, and F356W residual flux (within an elliptical aperture) with a correlation coefficient of $0.07$.

\subsection{Training the DeepMerge Algorithm on Simulated Galaxies} \label{sec:dm_training}
An alternative to using segmentation maps and morphology parameters is to train a machine learning algorithm directly on the image data. Here, we explore using the DeepMerge\footnote{\href{https://github.com/deepskies/deepmerge-public}{https://github.com/deepskies/deepmerge-public}} convolutional neural network \citep{cip2020} to identify mock CEERS mergers. Convolutional neural networks have special ``convolutional" layers which convolve the input image with a kernel, in order to extract features from the input image while preserving spatial information from the image. DeepMerge is a relatively simple network containing only three convolutional layers.

\begin{table}
\centering
\begin{tabular}{l|c|c|c}
\hline
\hline
Redshift Bin & $3 - 3.5$ & $3.5 - 4$  & $4 - 5$ \\
\hline
Total Number of Galaxies & 2039 & 2367 &  3105 \\
Number of Mergers & 668 & 839 & 1476 \\
\hline
\end{tabular}
\caption{Number of simulated CEERS mergers in each redshift bin after imposing $S/N>3$ and \texttt{Flag}$_{\texttt{Galfit}}$ $=2.0$.}
\label{tab:sim_counts}
\end{table}

As with the random forest analysis, we use the same mock CEERS sample (with S/N$_{F115W}$ $> 3$ and \texttt{Flag}$_{\texttt{Galfit}}$ $=2.0$). This time, we restrict our mock CEERS sample to three redshift bins at $3 < z  <3.5$, $3.5 < z  <4$, and $4 < z  <5$ (see Table \ref{tab:sim_counts}). We limit ourselves to these three redshift bins, since we will apply our networks to our observed CEERS galaxies for which visual classifications are only available for $z>3$. We cut out stamps of size 75 by 75 pixels of each galaxy. The time since last merger and time until next merger were again used to determine the ``labels" for DeepMerge.

\begin{figure}[t]
\plotone{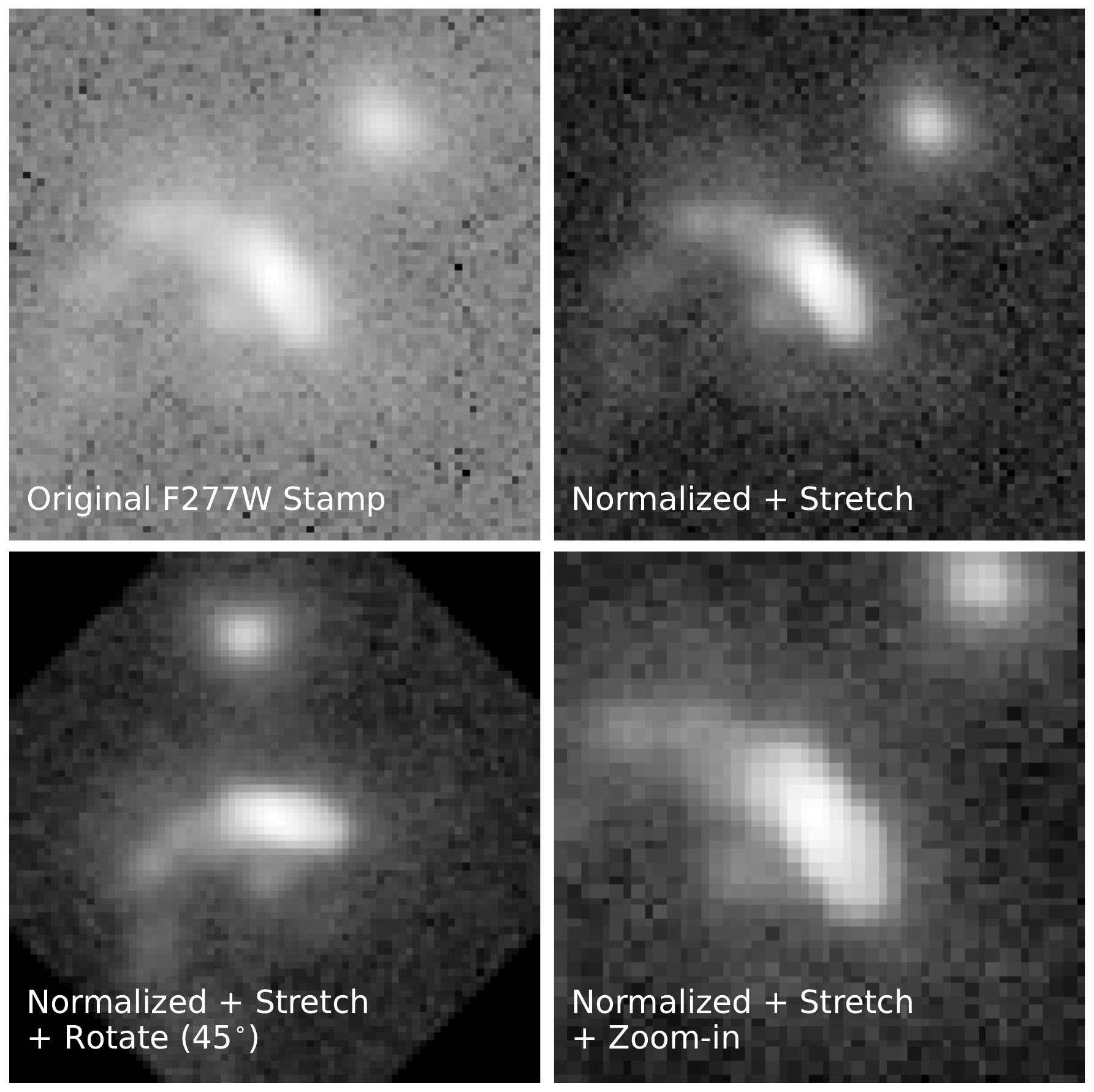}
\caption{Example of a normalized, stretched, and augmented $3 < z  <3.5$ galaxy merger. For this redshift bin, normalization was based on the F115W filter. After normalizing and stretching, the extended structures of this merger are more distinct from the background.
\label{fig:nn_augment}}
\end{figure}

We use stratified sampling to split our data into training, validation, and test sets. The ratio of the train/validation and test split was 0.18, which was smaller than the split used for the random forests in \cite{rose2023}, in order to divert more examples to the training set. The training and validation split was 0.3. Due to the low numbers of objects in each redshift bin, we employ augmentation to create a larger training set. This included flipping and rotating the images, as well as zooming in and zooming out while maintaining a consistent stamp size of 75 by 75 pixels. For each object, we provide DeepMerge with all six filters (F115W, F150W, F200W, F277W, F356W, F444W). We also normalize all images based on one of the six filters and then stretch the images using an arcsinh stretch to emphasize low surface brightness features. Figure \ref{fig:nn_augment} shows an example of normalizing, stretching, and augmenting a $3 < z  <3.5$ galaxy merger. Our data sets are imbalanced since there are more non-mergers than mergers. In order to help prevent our classifier from becoming biased toward the non-merger class, we maintain balanced classes during training by randomly undersampling the non-merger class, which decreases the size of the non-merger class. The validation and test sets are left unbalanced to better represent proportions of real-world data.

We start with the architecture of DeepMerge, and train using our data. We employ multiple strategies in order to properly train the DeepMerge networks. In addition to normalizing, augmenting, and balancing the training set, we also test:
\begin{enumerate}
\item Using different ratios for the split between the training, validation, and test sets.
\item Using different subsets of augmentations (e.g., only rotations).
\item Using only three filters, or only one rest-frame filter, rather than all six available filters.
\item Reducing the number of layers within the network, in order to reduce the complexity of the network.
\item Shifting the location of the dropout layers within the network. Dropout layers mask contributions from some neurons to the next layer in an effort to prevent overfitting.
\item Using rectified linear unit (ReLU) activation rather than sigmoid activation in the final layer. These activation functions determine how the result of the last layer will be mapped to a class probability.
\item Using different patience values for early stopping\footnote{\href{https://keras.io/api/callbacks/early\_stopping/}{https://keras.io/api/callbacks/early\_stopping/}}, which is a method that stops training once a given metric (in this case, validation loss) has stopped improving in order to combat overfitting \citep[e.g.,][]{Prechelt2012}.
\item Tracking the difference between training and validation accuracy, rather than validation loss, for early stopping.
\end{enumerate}
Most of the above strategies resulted in networks that performed very well in one class but very poorly in the other. We also attempted to narrow our merger definition to include only major mergers, but the small number of labeled mergers in the final data sets (with the $3 < z < 3.5$ bin having only 376 mergers) resulted in networks that classified all objects as non-mergers. We also attempted to include objects with any \texttt{Flag}$_{\texttt{Galfit}}$ (about $\sim10$ objects in each redshift bin), as this flag only pertains to the quality of the \texttt{Galapagos/Galfit} measurements which are not relevant here, but the inclusion of these objects also did not improve the performance of the networks. The reason for this may be that these objects were typically found in very crowded regions which may have caused DeepMerge to focus on the surrounding galaxies.

For all redshift bins, normalizing and arcsinh stretching the images resulted in improvements. We use all available filters. For the $3.5 < z < 4$ and $4 < z < 5$ redshift bins, we reduce the number of layers in DeepMerge to two sets of Convolution2D, BatchNormalization, MaxPooling2D, and Dropout layers rather than three sets. For the $3.0 < z < 3.5$ redshift bin, we also reduce the number of layers to two sets of Convolution2D, BatchNormalization, MaxPooling2D; additionally we use only one Dropout layer which we shifted to just before DeepMerge's Dense layers. We train for 500 epochs with a batch size of 32. For the $3 < z < 3.5$ redshift bin, we normalize with respect to the F115W filter; for the $3.5 < z < 4$ redshift bin, we normalize with respect to the F150W filter; and for the $4 < z < 5$ redshift bin, we again normalize with respect to the F150W filter. For each redshift bin, we perform the following augmentations on the training set: flip up-down, flip left-right, flip up-down and left-right, rotate at angles from $45^\circ$ to $270^\circ$, zoom-in by a factor of 1.5, and zoom-out by a factor of 0.75. For the $3 < z < 3.5$ bin, for early stopping, we set patience to 50 and monitor the validation loss; for the $3.5 < z < 4$ redshift bin, we set patience to 5 and monitor the ratio between validation and training accuracy so that training stops once this ratio is less than 0.8; for the $4 < z < 5$ redshift bin, we set patience to 50 and monitor the validation loss.

In \S \ref{sec:cnn_sim}, we describe the best outcome of training and testing the DeepMerge network for merger identification on the mock CEERS dataset at $3 < z < 5$. In \S \ref{sec:cnn_real}, we apply the DeepMerge to observed CEERS galaxies.

\section{Analysis} \label{sec:analysis}

\subsection{Visual Classification Results for Observed CEERS Galaxies} \label{sec:visclass}

We construct four possible ``merger" definitions based on the visual classifications. As described in \S \ref{sec:kartaltepe2023}, classifiers can select a main morphology class, an interaction class, and structure and quality flags. Our different merger definitions are motivated by these different categories:

\textit{Group 1 (G1)}: at least two out of the three classifiers assigned the Merger class or Interaction within segmentation map class or Interaction beyond segmentation map class.

\textit{Group 2 (G2)}: at least two out of the three classifiers assigned the Irregular main morphology class, since many merging galaxies appear irregular, although not all irregular galaxies are mergers.

\textit{Group 3 (G3)}: at least two out of the three classifiers assigned the tidal tails or double nuclei structure flags, which are merger signatures.

\textit{Group 4 (G4)}: one out of the three classifiers assigned the Merger class or Interaction within segmentation map class or Interaction beyond segmentation map class.

For groups 1, 2, and 3, we examine objects for which at least two out of three classifiers indicated that the object was potentially a merger, which means classifiers were overall more confident that these objects actually are mergers. For group 4, we examine objects where only only one classifier indicated that a given object had merger signatures. All objects from Group 1 will also be in this group, as well as additional objects for which classifiers were less confident about their merger status.

We start with the 798 objects that result from combining our morphology catalogs in \S \ref{sec:obs_morph_params}. Since we restricted our mock CEERS dataset to have S/N$_{F115W}$ $> 3$ and \texttt{Flag}$_{\texttt{Galfit}}$ $=2.0$, we do the same to our real CEERS dataset. Doing so results in a  final sample of 369 galaxies at $3 < z < 5$. Table \ref{tab:obs_counts} shows the total number of objects from this restricted real CEERS data set, as well as the number of mergers from each group, in our three redshift bins. In this paper, we focus on merger definitions using Groups 1, 2, and 4, since we found that most Group 3 objects are also in the other groups. Only two objects were assigned the tidal tails structure flag without also being assigned an interaction class by at least one person.

\begin{table}[t]
\centering
\begin{tabular}{l|c|c|c|c}
\hline
\hline
Redshift Bin & $3 - 3.5$ & $3.5 - 4$  & $4 - 5$ & $> 5$ \\
\hline
Total Number of Galaxies & 183 & 99 & 87 & 43 \\
Group 1 Mergers & 38 & 23 & 15 & 6 \\
Group 2 Mergers & 71 & 43 & 43 & 17 \\
Group 3 Mergers & 28 & 20 & 18 & 11 \\
Group 4 Mergers & 92 & 46 & 55 & 20 \\
\hline
\end{tabular}
\caption{Number of observed CEERS mergers in each redshift bin after imposing $S/N>3$ and \texttt{Flag}$_{\texttt{Galfit}}$ $=2.0$.}
\label{tab:obs_counts}
\end{table}

\begin{figure*}
\plotone{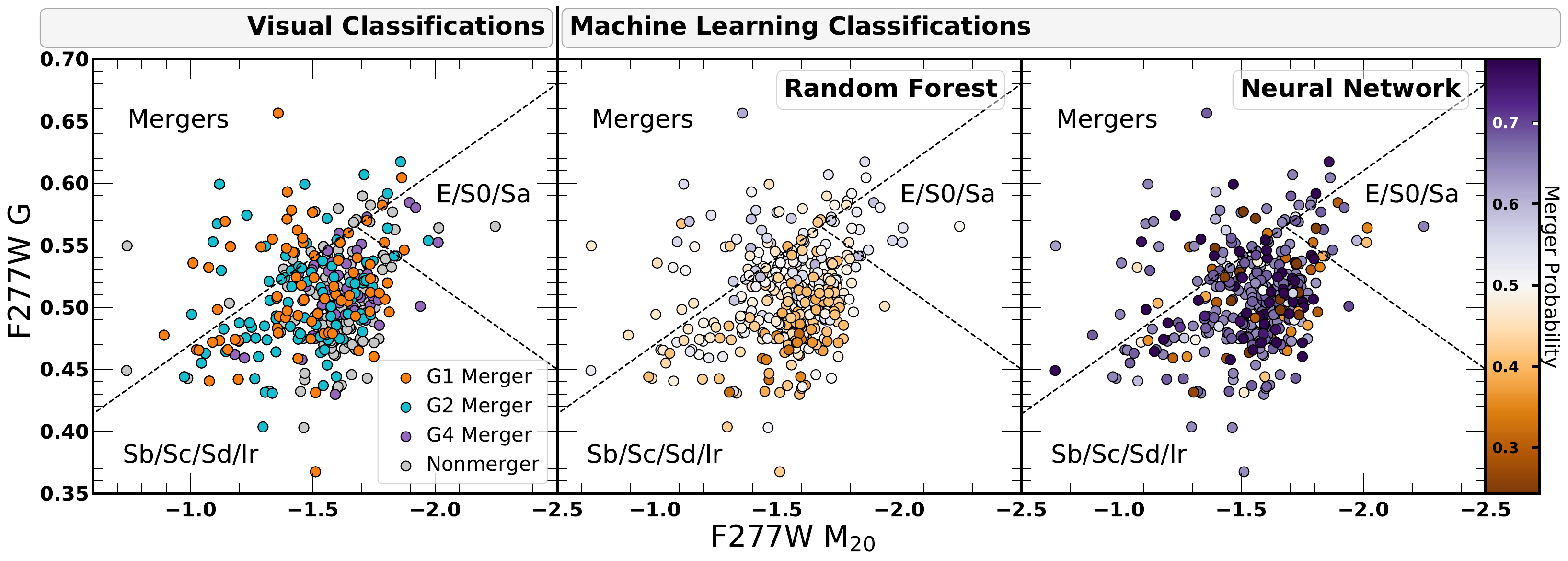}
\caption{F277W $G-M_{20}$ space for observed CEERS galaxies from $3 < z < 5$. In the \textit{\textbf{left panel}}, observed galaxies are color-coded by visual classifications as defined in \S \ref{sec:visclass} (\textit{Group 1 / G1}: at least two out of three volunteer classifiers assigned an interaction class; \textit{Group 2 / G2}: at least two out of three classifiers assigned the Irregular morphology class; \textit{Group 4 / G4}: one of out three classifiers assigned an interaction class). In the \textit{\textbf{middle panel}} and \textit{\textbf{right panel}}, galaxies are color-coded by the merger probability output by the random forest or neural network, respectively.
\label{fig:g_m20}}
\end{figure*}

\subsection{Classical Techniques: Performance on Observed CEERS Galaxies} \label{sec:ch3_classical}
Past studies have often used classical techniques such as $G$ vs. $M_{20}$ \citep{lotz2004, lotz2008} to identify mergers. In such plots, the merger discriminating line is 
\begin{equation} \label{eq:gm20}
    G > -0.14 M_{20} + 0.33,
\end{equation}
defined by \cite{lotz2008}. 
The first panel of Figure \ref{fig:g_m20} shows $G$ vs. $M_{20}$ for the observed CEERS dataset, color-coded by visual classification (Groups 1, 2, and 4 have been abbreviated as G1, G2, and G4 in the legend). $G$ vs.~$M_{20}$ does not appear to effectively separate visually classified mergers and non-mergers, which is to be expected since $G-M_{20}$ is not sensitive to all stages of a merger. The fraction of correctly classified mergers in each group, based on the merger discriminating line, is only $27.6\%$ for Group 1, $21.7\%$ for Group 2, and $19.2\%$ for Group 4. Of the objects in the merger region above the merger discriminating line, only $35.6\%$ are actually Group 1 mergers, $57.6\%$ are actually Group 2 mergers, and $62.7\%$ are actually Group 4 mergers. Most objects tend to fall in the late-type region of the plot. This performance of the $G-M_{20}$ method is similar to that in Figure 6 of \cite{rose2023} for the simulated CEERS data, where only $19-23\%$ of the simulated mergers were correctly classified, and only $4-50\%$ of predicted mergers were actually mergers.

$G-M_{20}$ has been shown to be unable to select a pure sample of mergers in other studies. For example, \cite{con2008} measured $G$ and $M_{20}$ for visually-classified HST UDF galaxies in several redshift bins from $0.4 < z < 3$ and find that, besides peculiars/mergers, most other galaxy types were actually also found above the merger discriminating line. In their last redshift bin of $2.5 < z < 3$, they find that $\sim93\%$ of peculiars are found in the merger region, and of all objects in the merger region, $\sim67\%$ are peculiars. \cite{pearson2019_gm20} plot $G-M_{20}$ for $0.002 < z < 0.15$ KiDS-GAMA galaxies with visual classifications from Galaxy Zoo. They report no specific numbers for the fraction of mergers selected by $G-M_{20}$, but find that most mergers and many non-mergers appear in the merger region. However, they then also include visual classifications from \cite{darg1,darg2}, and find that most \cite{darg1,darg2} mergers lie in the non-merger region. \cite{kartaltepe2010} plot $G-M_{20}$ for $0.01 < z < 3.5$ visually-classified COSMOS galaxies and also do not report specific numbers, but find that most (major and minor) mergers appear in the non-merger region. They explored several automated methods besides $G-M_{20}$ and find that no technique is sensitive to all merger stages. \cite{sny2019} determine $G-M_{20}$ for simulated Illustris-1 galaxies, with mergers defined by a timescale of $\pm250$ Myr, and also do not report specific numbers, but found that diagnostics including $G-M_{20}$ select a highly incomplete sample of mergers, as most mergers appeared in the non-merger region.

The second and third panels show the merger classifications predicted by the random forest (\S \ref{sec:rf_real}) and the DeepMerge network (\S \ref{sec:cnn_real}), respectively. These panels are both color-coded by the color bar at the far right, which represents the merger probability output by either the random forest or neural network. Probabilities higher than 0.5 mean the object was classified as a merger (and colored purple or light purple). Probabilities lower than 0.5 mean the object was classified as a non-merger (and colored orange or yellow). See \S \ref{sec:rf_nn_gm20} for further discussion of these two panels.

\subsection{Random Forests: Performance on Simulated CEERS Galaxies} \label{sec:rf_sim}

Here we summarize the results from training random forests on mock CEERS galaxies from IllustrisTNG in \cite{rose2023}. Across all redshift bins, the forests accurately classified about $\sim60\%$ of merging and non-merging galaxies (according to our merger definition of $\pm250$ Myr). The forests performed better than both a random classifier and classical techniques. For lower redshift bins, rest-frame asymmetry features were more important and for higher redshift bins, rest-frame bulge and clump features were more important to the forests for merger identification. The feature importances were calculated by the random forest and accessed via the \texttt{feature\_importances\_} attribute. The forests misclassified non-mergers that had segmentation maps contaminated with emission from nearby or background galaxies or had merger signature persisting from outside the chosen merger time frame. The forests also misclassified mergers that appeared visually undisturbed. Finally, the merger fraction and merger rate calculated using random-forest-selected mergers were underestimated compared to previous works that used Illustris-1.

\begin{figure*}
\plottwo{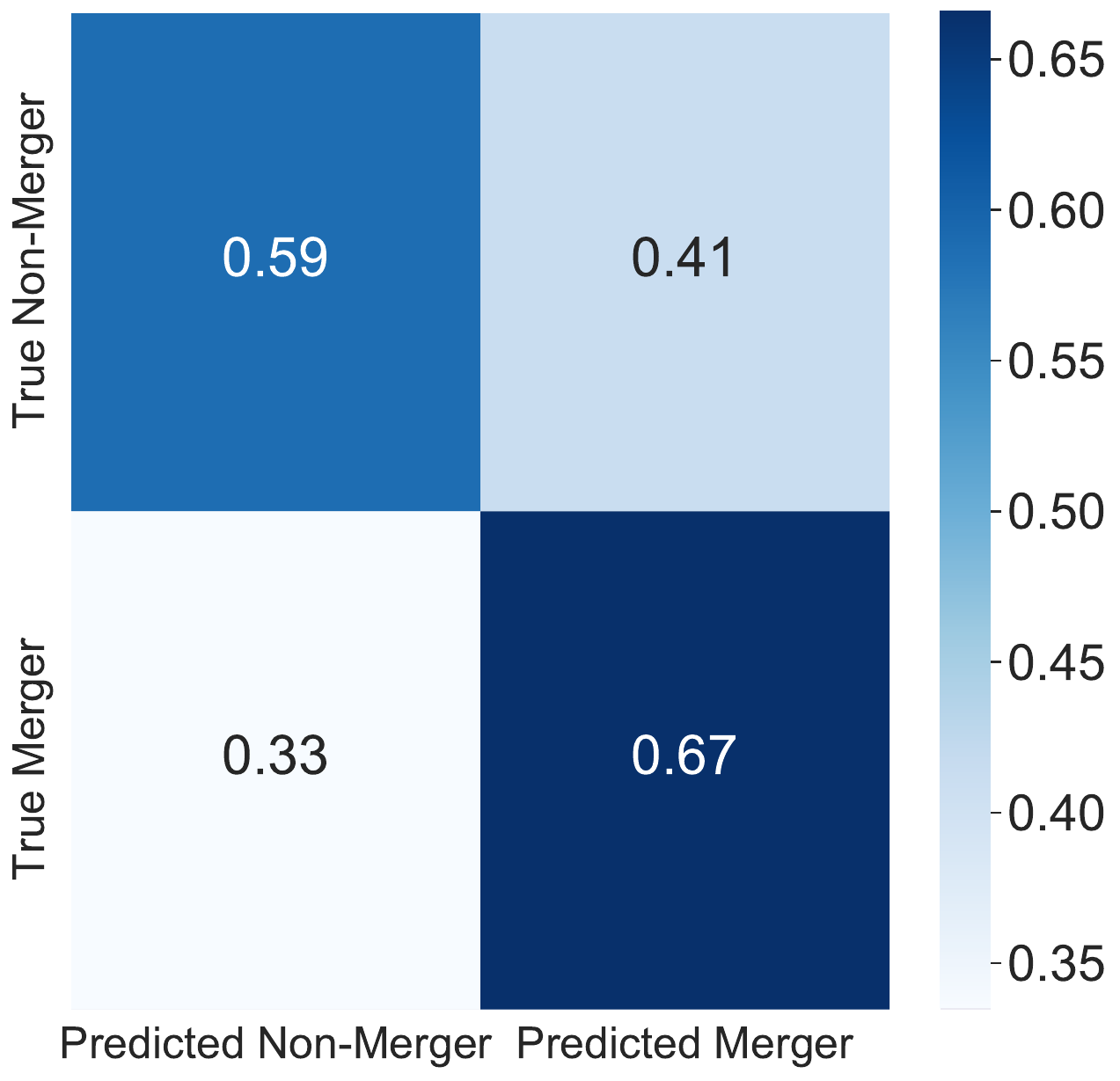}{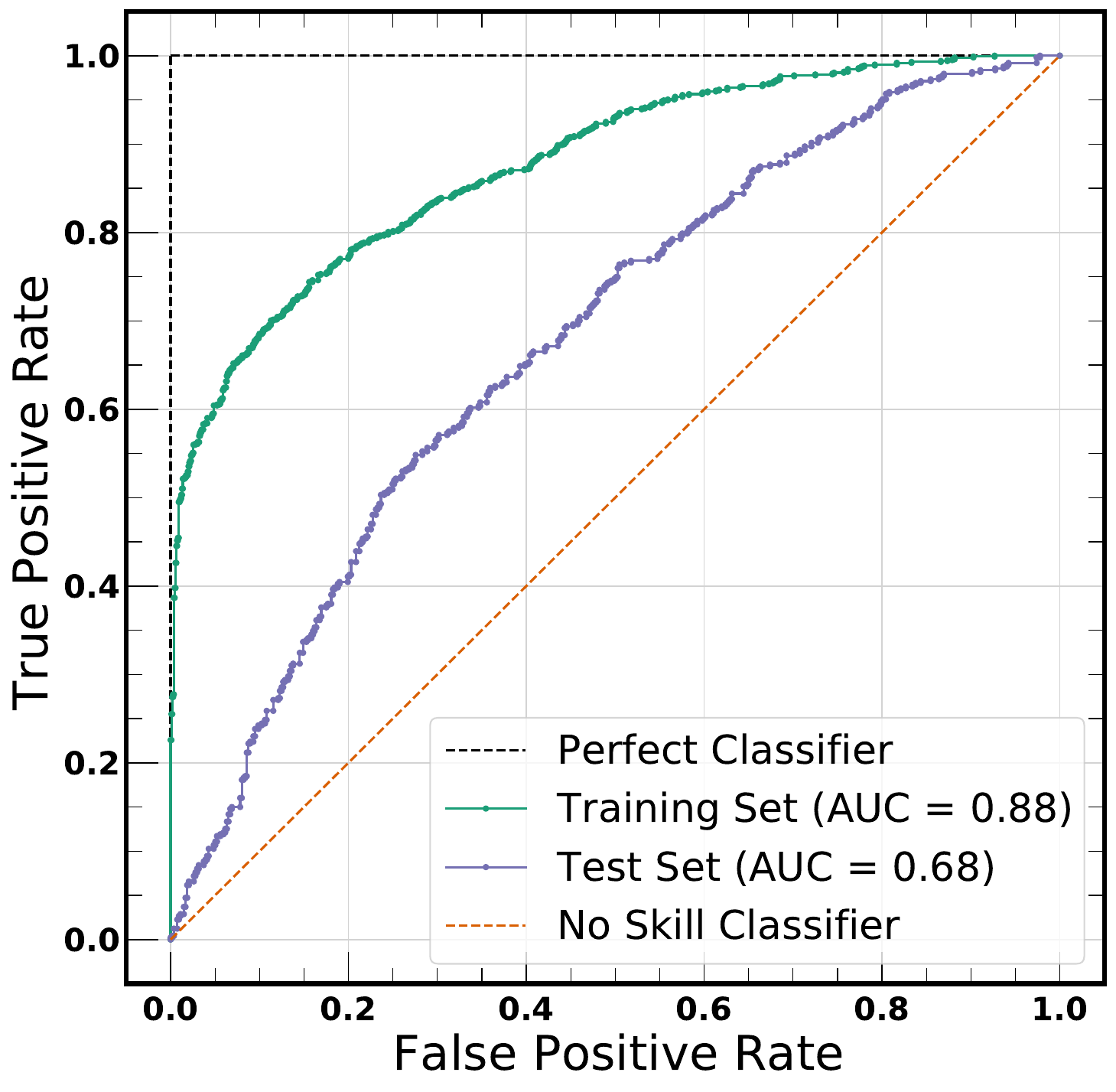}
\caption{\textit{\textbf{Left}}: Random forest confusion matrix for $4 < z < 5$ objects from the
test set. The diagonal shows the fraction of objects correctly classified for each class. \textit{\textbf{Right}}: ROC curve for the $4 < z < 5$ random forest. The training set (\textit{green}) and test set (\textit{purple}) curves lie in the region of ``good" classifiers (between the perfect (\textit{black}) and no skill (\textit{red}) classifiers).
\label{fig:rf_4_5}}
\end{figure*}

In this work, we additionally trained a random forest for $z = 4 - 5$. This forest correctly classified $59\%$ of non-merging galaxies and $67\%$ of merging galaxies. The confusion matrix and receiver operating characteristic (ROC) curve are shown in Figure \ref{fig:rf_4_5}, which indicate similar performance to the forests in \cite{rose2023}.

\subsection{DeepMerge: Performance on Simulated CEERS Galaxies} \label{sec:cnn_sim}

\begin{figure*}
\minipage{\textwidth}%
  
  \includegraphics[width=0.33\linewidth]{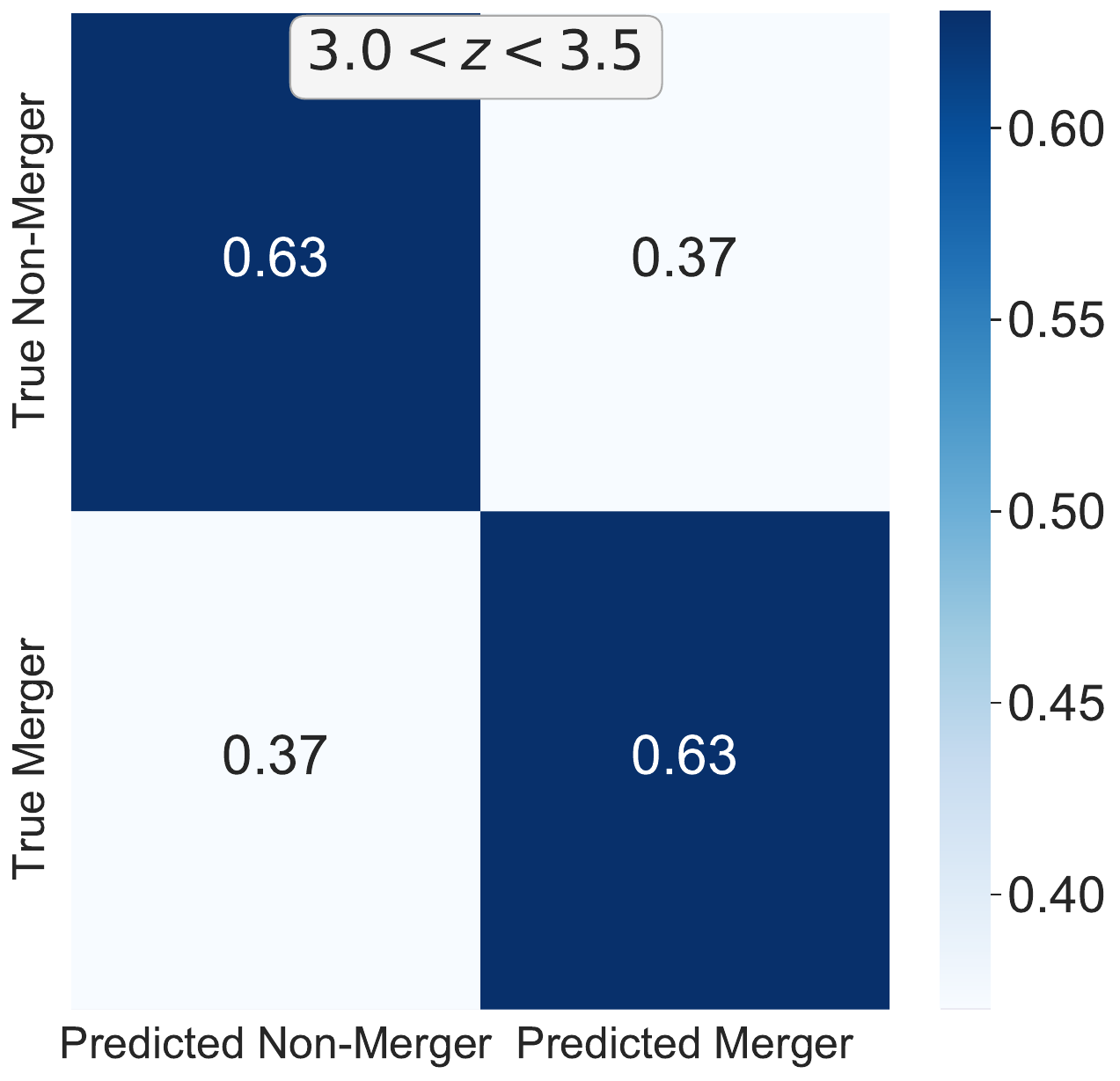}
  \includegraphics[width=0.33\linewidth]{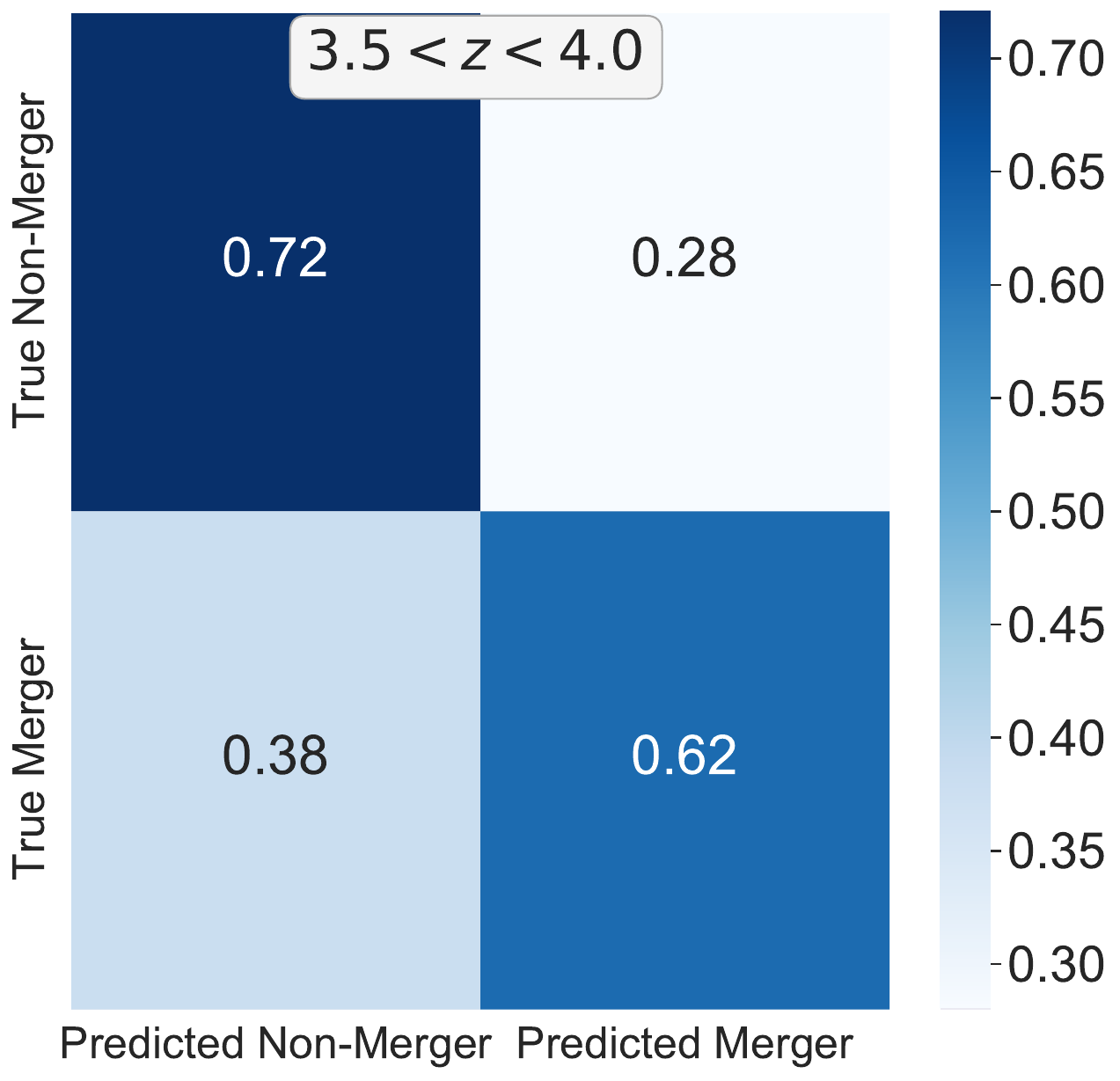}
  \includegraphics[width=0.33\linewidth]{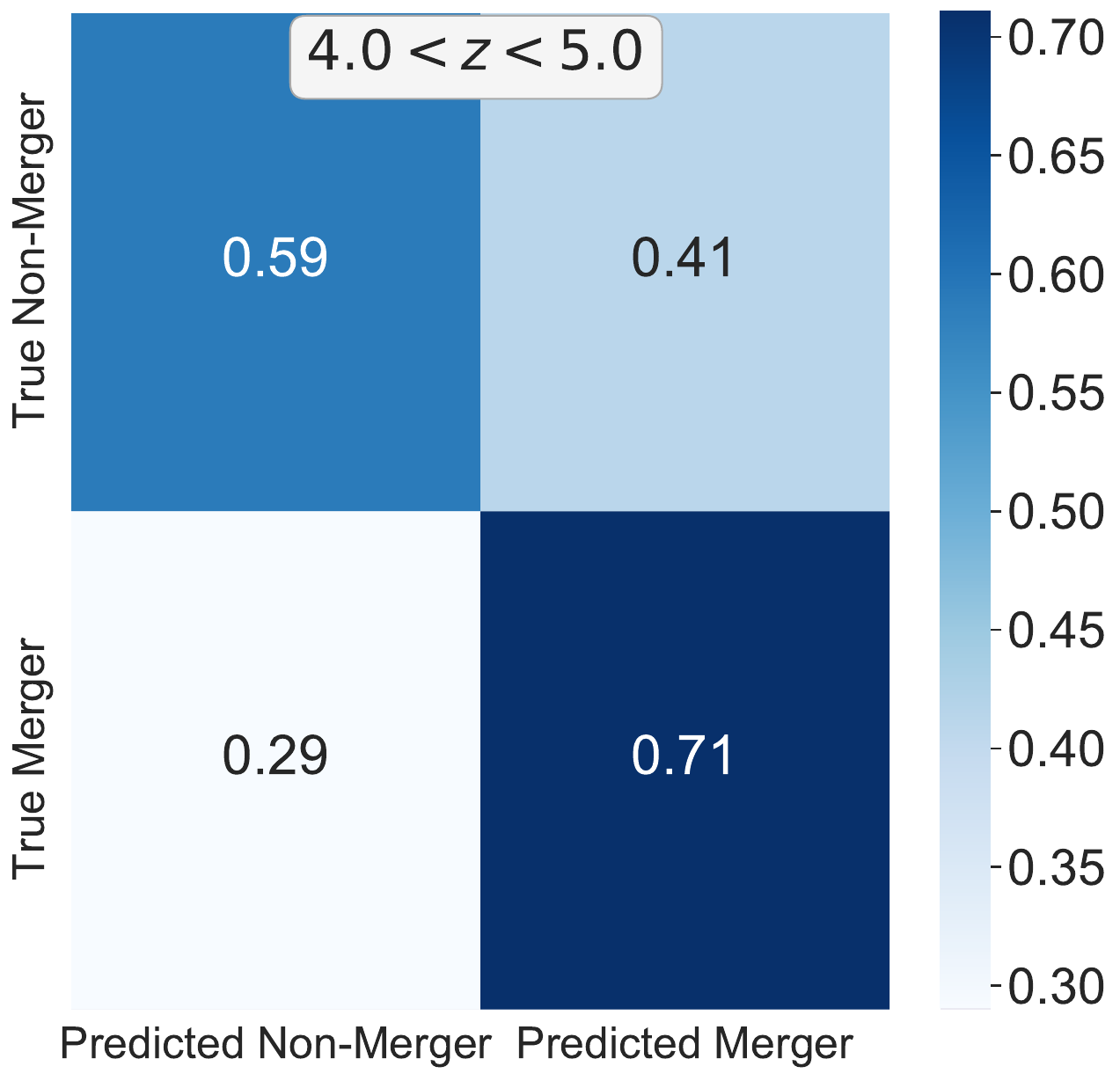}
\endminipage\hfill
\vskip2ex
\minipage{\textwidth}%
  
  \includegraphics[width=0.33\linewidth]{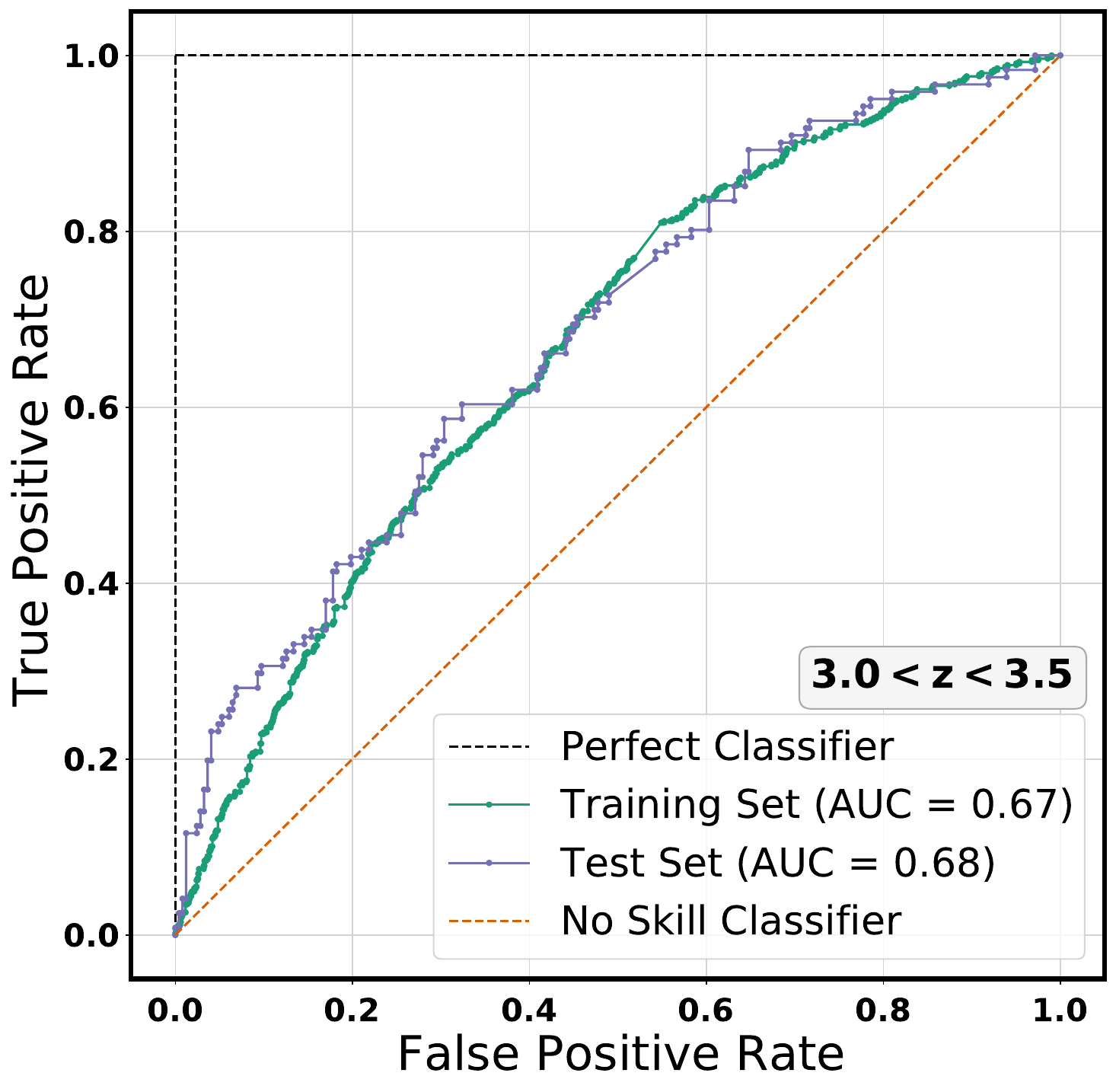}
  \includegraphics[width=0.33\linewidth]{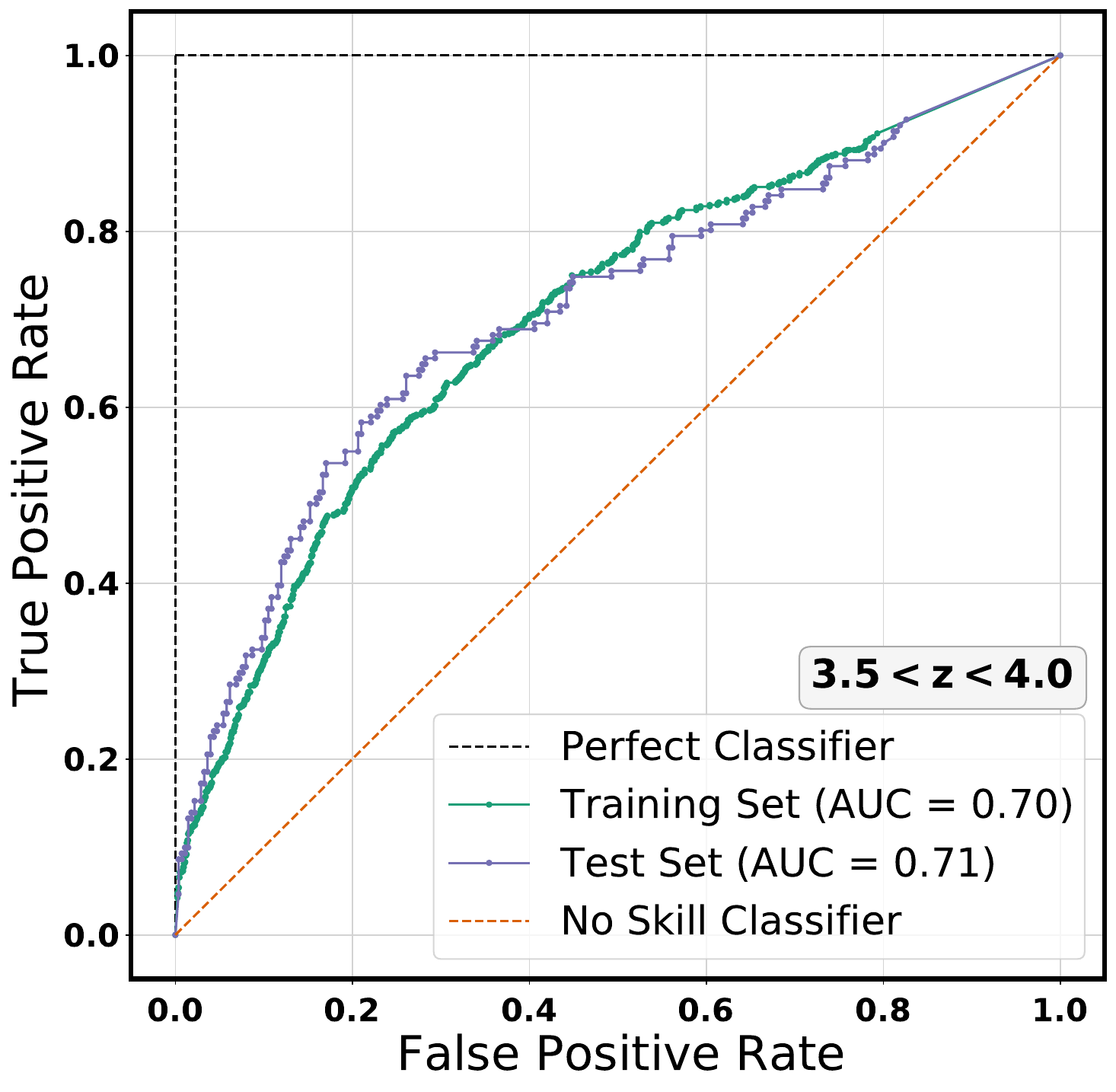}
  \includegraphics[width=0.33\linewidth]{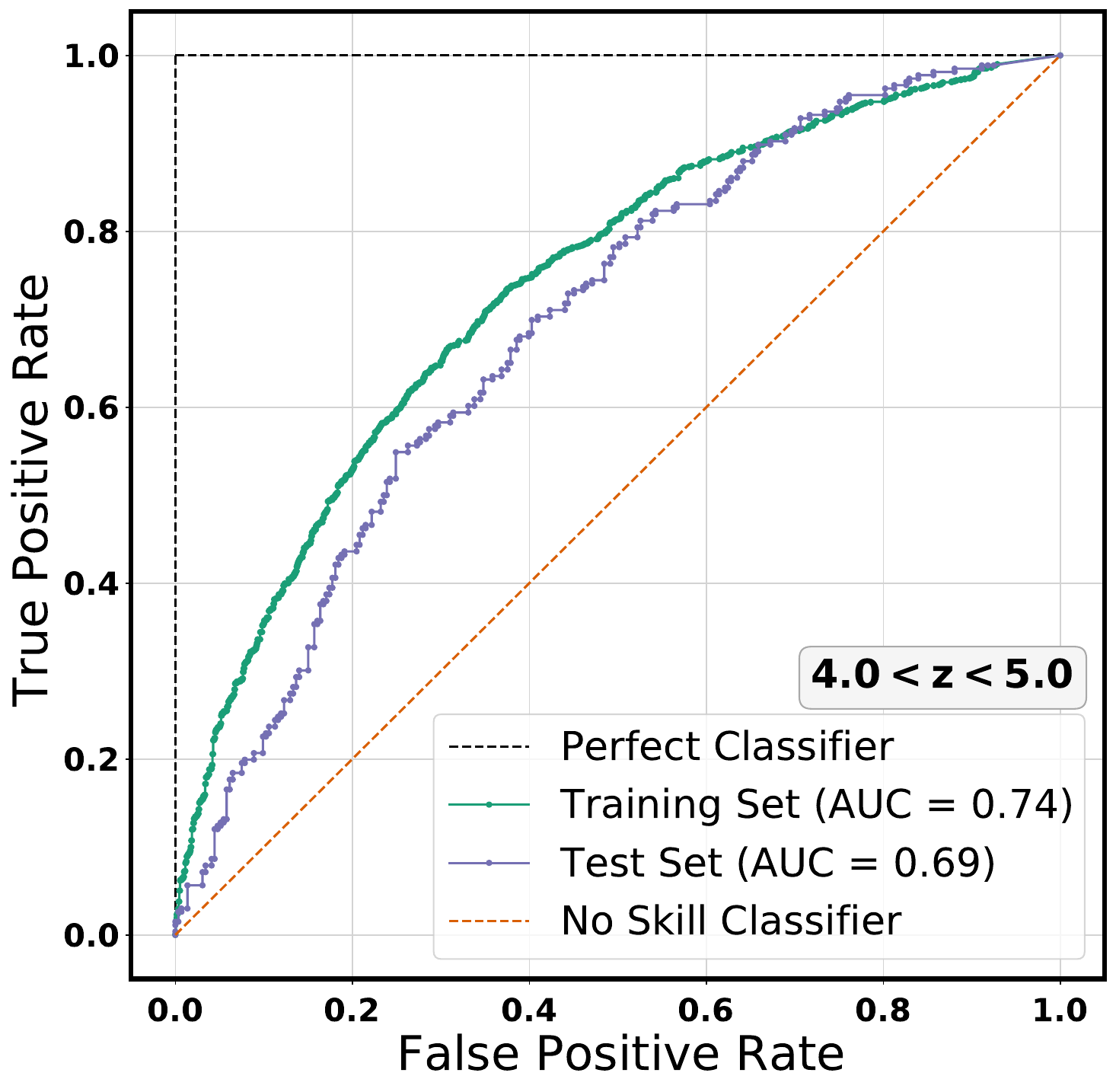}
  
\endminipage
\caption{\textit{\textbf{Top}}: DeepMerge confusion matrices for the simulated CEERS test set, for each redshift bin. \textit{\textbf{Bottom}}: Corresponding ROC curves for the simulated CEERS test set, for each redshift bin. The training set (\textit{green}) and test set (\textit{purple}) curves lie in the region of ``good” classifiers (between the perfect (\textit{black}) and no skill (\textit{red}) classifiers).} \label{fig:dm_sim_results}
\end{figure*}

Figure \ref{fig:dm_sim_results} shows the confusion matrices and ROC curves resulting from applying our trained DeepMerge models to the simulated CEERS test set. Our DeepMerge networks correctly classify $\sim 60 - 70\%$ of merging and non-merging galaxies. The ROC curves show that the networks perform better than a random classifier. As in \cite{rose2023}, we categorize the test set results into four classes:
\begin{itemize}
\item True Positives (TP): the number of true mergers correctly classified by the random forest.
\item False Positives (FP): the number of non-mergers incorrectly classified as mergers.
\item True Negatives (TN): the number of correctly classified non-mergers.
\item False Negatives (FN): the number of true mergers incorrectly classified as non-mergers.
\end{itemize}

There are several metrics used to quantify the performance of the neural networks:
\begin{itemize}
\item True Positive Rate (TPR) -- also known as \textit{recall} or \textit{completeness}:
    \begin{equation}
        TPR = \textrm{recall} = \frac{TP}{TP + FN}.
    \end{equation}
\item False Positive Rate (FPR) -- also known as \textit{fall out}:
    \begin{equation}
        FPR = \frac{FP}{FP + TN}.
    \end{equation}
\item Positive Predictive Value (PPV) -- also known as \textit{precision}:
    \begin{equation}
        PPV = \textrm{precision} = \frac{TP}{TP + FP}.
    \end{equation}
\item F1 Score -- the harmonic mean of precision (P) and recall (R):
    \begin{equation}
    F_{1} = 2\frac{P \times R}{P + R} = \frac{TP}{TP + \frac{1}{2} (FP + FN)}.
    \end{equation}
\end{itemize}
Since our simulated CEERS datasets are imbalanced, the F1 score is a more useful metric than accuracy for comparing the performance of our models. Table \ref{tab:sim_f1score} compares the F1 scores of the random forests and DeepMerge networks for the simulated data. Based on the F1 scores, the DeepMerge networks technically performed better than the random forests in each redshift bin, particularly the $3.5 < z < 4$ bin.

\begin{table}
\centering
\begin{tabular}{l|c|c|c}
\hline
\hline
Redshift Bin & $3 - 3.5$ & $3.5 - 4$  & $4 - 5$ \\
\hline
RF Non-mergers & 0.67 & 0.65 & 0.62 \\
NN Non-mergers & 0.70 & 0.75 & 0.64 \\
\hline
RF Mergers & 0.52 & 0.54 & 0.63 \\
NN Mergers & 0.53 & 0.59 & 0.66 \\
\hline
\end{tabular}
\caption{Comparison of F1 scores for the random forests (RF) and DeepMerge neural networks (NN) on the simulated CEERS test sets.}
\label{tab:sim_f1score}
\end{table}

\begin{figure}
\plotone{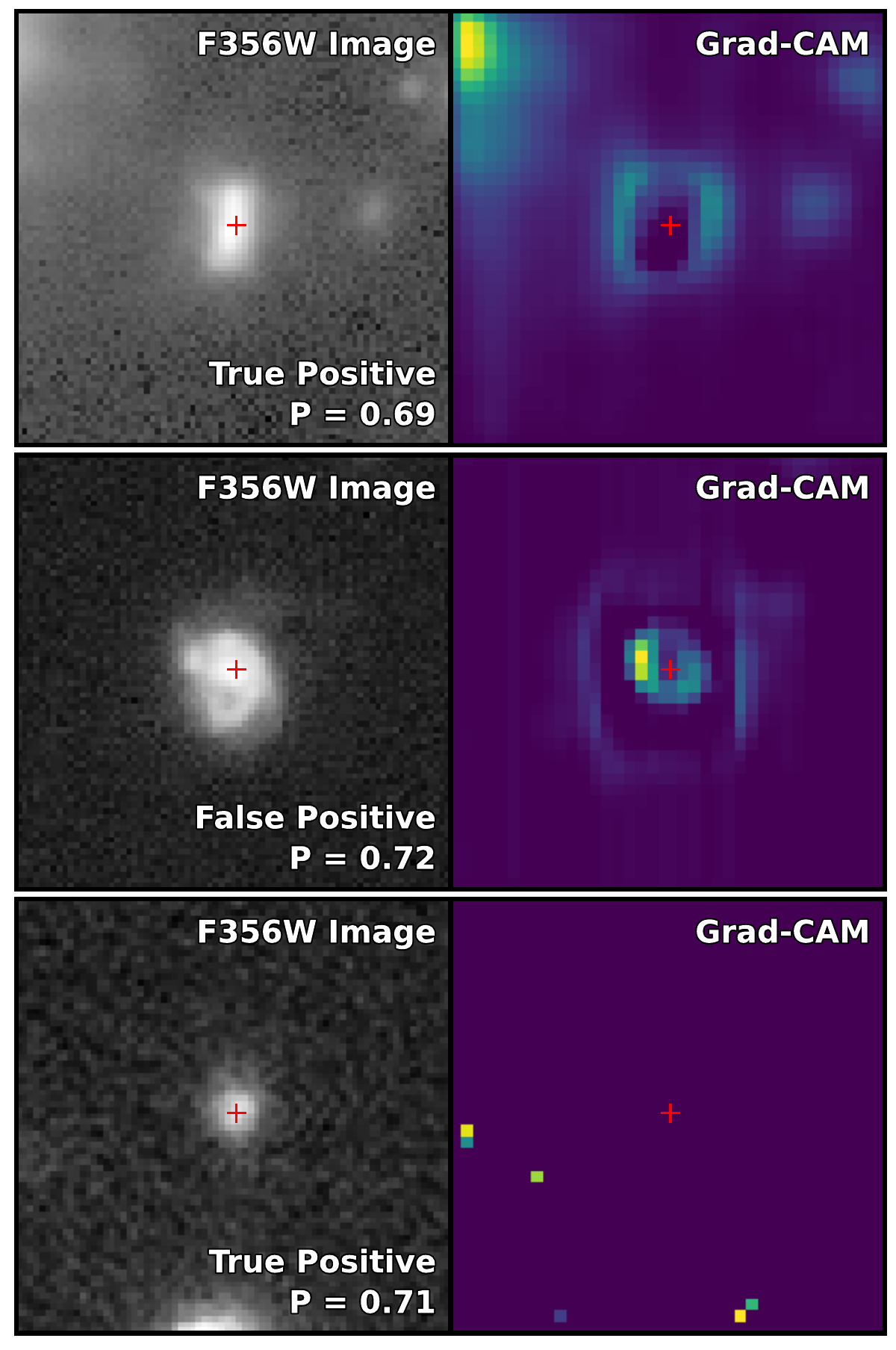}
\caption{Examples of simulated CEERS galaxies and their corresponding Grad-CAMs from the single-filter DeepMerge test using F356W images. The bottom left corner of the F356W image stamps shows the categorization (e.g., true positive, false positive, etc.) of this object and the DeepMerge merger probability. The red crosses show the location of the center of each image.
\label{fig:gradcam}}
\end{figure}

As in \cite{cip2020}, we investigate the use of Gradient-weighted Class Activation Mapping (Grad-CAM) maps \citep{selva2020}. These mappings, calculated from the gradient of the predicted class of the input image with respect to the activations in the last convolutional layer, allow us to validate that the network is ``looking" at the correct regions in the input image. However, for our complex data with six filters, the Grad-CAMs were not directly interpretable and showed a six-by-six pattern of activations all around the input image. Therefore, we investigate Grad-CAMs from one of the previous tests described in \S \ref{sec:dm_training}, where we use only one rest-frame filter rather than all six. Figure \ref{fig:gradcam} shows some examples of Grad-CAMs from the $z=4-5$ bin trained using only the F356W filter. For this test, the network correctly classified $62\%$ of simulated CEERS non-mergers and $53\%$ of simulated CEERS mergers. As shown in Figure \ref{fig:gradcam}, some Grad-CAMs from this F356W filter test show that the network does in fact activate around the edges of the central galaxies and around the edges of neighbors as we would expect, but other Grad-CAMs show only activations in a few random pixels. The distributions of the F356W magnitude and signal-to-noise per pixel indicate that objects with these ``bad" Grad-CAMs are not necessarily more likely to have a fainter magnitude and lower S/N. The peak magnitude for objects with ``bad" Grad-CAMs is $28.8 \pm 1.2$, while the peak magnitude for objects with ``good" Grad-CAMs is $28.1 \pm 1.5$. Of objects with ``bad" Grad-CAMs, $55.7\%$ were correctly classified (either true positives or true negatives) and $44.4\%$ were incorrectly classified (either false positive or false negative). Of the objects with ``good" Grad-CAMs, $57.6\%$ were correctly classified and $42.4\%$ were incorrectly classified.

\subsection{Random Forests: Performance on Observed CEERS Galaxies} \label{sec:rf_real}

\begin{figure}
\plotone{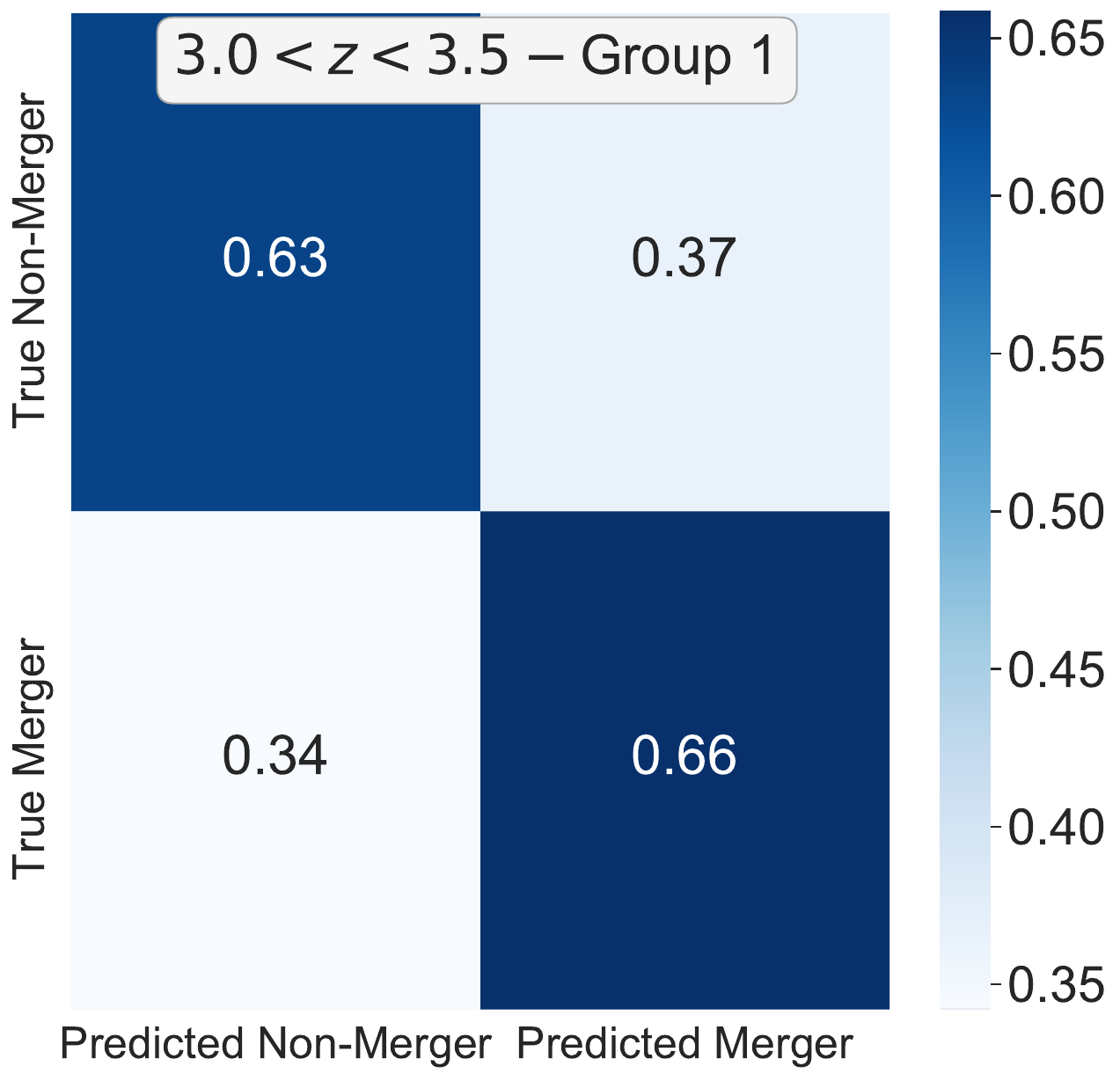}
\caption{Random forest confusion matrix for observed CEERS galaxies at $3 < z < 3.5$.
\label{fig:rfobsconmat}}
\end{figure}

After training on the simulated images, we apply the random forest algorithm to observed CEERS galaxies, selected in \S \ref{sec:visclass}, at the following redshift bins: $3 < z < 3.5$, $3.5 < z < 4$, and $4 < z < 5$. Figure \ref{fig:rfobsconmat} shows the confusion matrix for the $3 < z < 3.5$ redshift bin for Group 1 (Figure \ref{fig:rf_real_cm} in the Appendix shows all the Group 1, 2, and 4 confusion matrices for each redshift bin). Here, we use the visual classifications as the true merger labels for the comparison. Since the visual classifications are subjective and volunteer classifiers may misclassify objects or disagree on classifications, the true merger labels may not be correct. In summary, the forests generally did not perform as well on the observed galaxies as they did on the simulated galaxies in the test sets. The only dataset where the random forest performed better than on the test set was Group 1 at $3 < z < 3.5$, where the forest correctly classified $63\%$ of non-mergers and $66\%$ of mergers. The forests performed the worst in the highest redshift bin, where only $25\% - 30\%$ of mergers were correctly classified. Table \ref{tab:obs_metrics} shows how the precision, recall, and F1-score change as a function of redshift for these groups.

\begin{figure}
\plotone{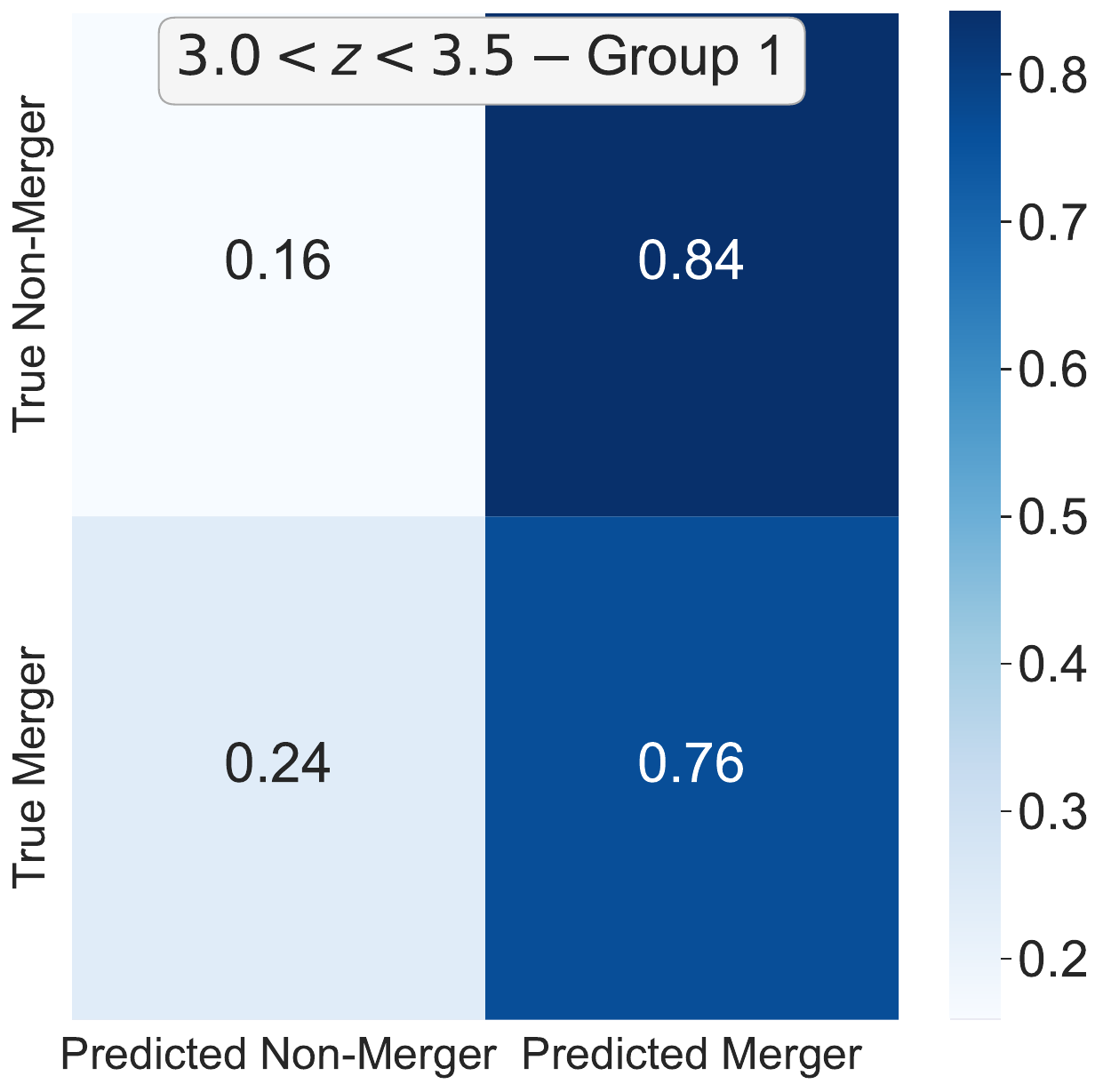}
\caption{DeepMerge neural network confusion matrix for observed CEERS galaxies at $3 < z < 3.5$.
\label{fig:nnobsconmat}}
\end{figure}

\begin{table*}
\centering
\begin{tabular}{l l l|cc|cc|cc}
\hline
\hline
& & & \multicolumn{2}{c|}{$3.0 < z < 3.5$} & \multicolumn{2}{c|}{$3.5 < z < 4$}  & \multicolumn{2}{c}{$4 < z < 5$} \\
\hline
& & & Non-merger & Merger & Non-merger & Merger & Non-merger & Merger \\
\hline
\hline
& & Precision & 0.88 & 0.32 & 0.71 & 0.19 & 0.84 & 0.24 \\
& Group 1 & Recall & 0.63 & 0.66 & 0.38 & 0.48 & 0.82 & 0.27 \\
& & F1-score & 0.74 & 0.43 & 0.50 & 0.27 & 0.83 & 0.25  \\
\cline{2-9}
& & Precision & 0.64 & 0.42 & 0.51 & 0.40 & 0.57 & 0.76 \\
Random Forest & Group 2& Recall & 0.60 & 0.46 & 0.38 & 0.53 & 0.91 & 0.30 \\
& & F1-score & 0.62 & 0.44 & 0.43 & 0.46 & 0.70 & 0.43  \\
\cline{2-9}
& & Precision & 0.59 & 0.63 & 0.56 & 0.48 & 0.41 & 0.82 \\
& Group 4 & Recall & 0.68 & 0.53 & 0.43 & 0.61 & 0.91 & 0.25 \\
& & F1-score & 0.63 & 0.58 & 0.49 & 0.54 & 0.57 & 0.39  \\
\hline
\hline
& & Precision & 0.72 & 0.19 & 0.64 & 0.22 & 0.73 & 0.14 \\
& Group 1 & Recall & 0.16 & 0.76 & 0.09 & 0.83 & 0.22 & 0.60 \\
& & F1-score & 0.26 & 0.31 & 0.16 & 0.34 & 0.34 & 0.22 \\
\cline{2-9}
& & Precision & 0.53 & 0.37 & 0.45 & 0.42 & 0.41 & 0.46\\
DeepMerge Network & Group 2& Recall & 0.15 & 0.79 & 0.09 & 0.86 & 0.20 & 0.70 \\
& & F1-score & 0.24 & 0.50 & 0.15 & 0.56 & 0.27 & 0.56 \\
\cline{2-9}
& & Precision & 0.44 & 0.49 & 0.27 & 0.43 & 0.23 & 0.58 \\
& Group 4 & Recall & 0.15 & 0.80 & 0.06 & 0.83 & 0.16 & 0.69\\
& & F1-score & 0.23 & 0.61 & 0.09 & 0.57 & 0.19 & 0.63 \\
\hline
\end{tabular}
\caption{Precision, recall, and F1-score for merger identification of observed CEERS galaxies, for both random forests and DeepMerge. \textit{Group 1 (G1) mergers} are observed galaxies for which at least two out of three volunteer classifiers assigned an interaction class. \textit{Group 2 (G2) mergers} are observed galaxies for which at least two out of three volunteer classifiers assigned the Irregular morphology class. \textit{Group 4 (G4) mergers} are observed galaxies for which at least one out of three volunteer classifiers assigned an interaction class.}
\label{tab:obs_metrics}
\end{table*}

\subsection{DeepMerge: Performance on Observed CEERS Galaxies} \label{sec:cnn_real}

After training on the simulated images, we apply our DeepMerge networks to observed CEERS galaxies at the same redshift bins of $3 < z < 3.5$, $3.5 < z < 4$, and $4 < z < 5$. Figure \ref{fig:nnobsconmat} shows the confusion matrix for the $3 < z < 3.5$ redshift bin for Group 1 (Figure \ref{fig:nn_real_cm} in the Appendix shows all the Group 1, 2, and 4 confusion matrices for each redshift bin). In summary, the networks did not perform as well on the observed galaxies as they did on the simulated galaxies in the test sets. At $3 < z < 3.5$ and $4 < z < 5$ the networks classified $\sim 15-20\%$ of non-mergers correctly and $\sim 70-80\%$ of mergers correctly across all visual classification groups. The networks performed the worst at $3.5 < z < 4$, with only $6-9\%$ of non-mergers correctly classified and $\sim85\%$ of mergers correctly classified for all visual classification groups. Table \ref{tab:obs_metrics} shows how the precision, recall, and F1-score change as a function of redshift for these groups.

Based on the confusion matrices, the DeepMerge networks tend to incorrectly classify more objects as mergers than the random forests. In addition to the strategies employed in \S \ref{sec:dm_training}, we additionally tried the following strategies during training to prevent underfitting and improve performance on the observed CEERS data set:
\begin{enumerate}
\item Giving DeepMerge only one rest-frame filter to train on, or summing the six filters together into one image, in order to reduce the complexity of the problem.
\item Reducing the number of layers in the network, in order to reduce the complexity of the network.
\item Adjusting the probability threshold for merger classification (default 0.5), such that only objects with merger probabilities higher than $0.55$, $0.6$, or $0.65$ were identified as mergers.
\end{enumerate}
The results of these tests did not significantly improve performance on the observed CEERS data set. While some tests did result in confusion matrices that were slightly more balanced, the networks still either classified almost all objects as mergers or almost all objects as non-mergers.

\section{Discussion} \label{sec:discuss}

\subsection{Classical Techniques: Observed
CEERS Galaxies with Machine Learning Classifications}\label{sec:rf_nn_gm20}

The second and third panels of Figure \ref{fig:g_m20} show $G$ vs. $M_{20}$ for the observed CEERS dataset, color-coded by the merger probabilities predicted by the random forests and the DeepMerge network, respectively. Objects were classified as mergers if the merger probability was greater than 0.5, and classified as non-merger if the merger probability was less than 0.5. As we saw in \S\ref{sec:ch3_classical}, the fraction of correctly classified mergers in each group, based on the merger discriminating line, ranged from only $19.2 - 27.6\%$. Thus, the machine learning algorithms were able to correctly classify more mergers than the $G-M_{20}$ method, and Figure \ref{fig:g_m20} shows that many of these mergers are below the $G-M_{20}$ merger discriminating line as expected given that $G-M_{20}$ only selects mergers at a narrow portion of the merger phase. Therefore the algorithms are able to identify mergers that the $G-M_{20}$ method misses due $G-M_{20}$'s limited sensitivity. A visual inspection of the second panel of Figure \ref{fig:g_m20} reveals a potential trend where objects with very low probabilities from the random forest are far below the merger discriminating line while objects with higher probabilities are closer to or above the line. A visual inspection of the third panel of Figure \ref{fig:g_m20} shows that most galaxies were classified as mergers with relatively high probabilities, and there appears to be no clear trend in relation to the merger discriminating line.

\subsection{Feature Importance}

In \cite{rose2023}, we calculated the relative importance of the features (morphology parameters) given to the random forests, which shows how useful a given feature is for identifying simulated mergers. We found that at lower redshift, bluer filters and asymmetry features were more useful, while at higher redshift, redder filters and bulge and clump features were more useful. Here we calculate feature importance for the $4 < z < 5$ random forest for simulated galaxies, which are (in order of importance):
\begin{enumerate}
    \item F277W multimode ($M$)
    \item F444W moment of light ($M_{20}$)
    \item F356W Gini ($G$)
    \item F444W concentration ($C$)
    \item F356W concentration ($C$)
\end{enumerate}
This continues the trend seen in Figure 11 in \cite{rose2023} where redder filter and bulge and clump features are more important at higher redshift.

Of the most correlated features identified by the Spearman correlation in \S\ref{sec:rf_training}, we see that only one of those features was present in the top five most important features calculated by the random forest, indicating that the Spearman correlation is not necessarily a good predictor of the random forest results.

It is important to note that morphology parameters are known to be sensitive to different merger stages. For example, \cite{snylotz2015} find that, when simulating a galaxy as observed with HST from $z \sim 3$ to $z \sim 1$ (with a merger at $z = 1.6$), the $G-M_{20}$ merger statistic is sensitive to the early phases of the merger, while the $MID$ statistics are more sensitive after coalescence. Future studies with larger datasets could benefit from investigating random forest feature importance as a function of merger stage, in addition to redshift.

\subsection{Galaxy Categorization Results}

\begin{figure*}[t]
\includegraphics[scale=0.6]{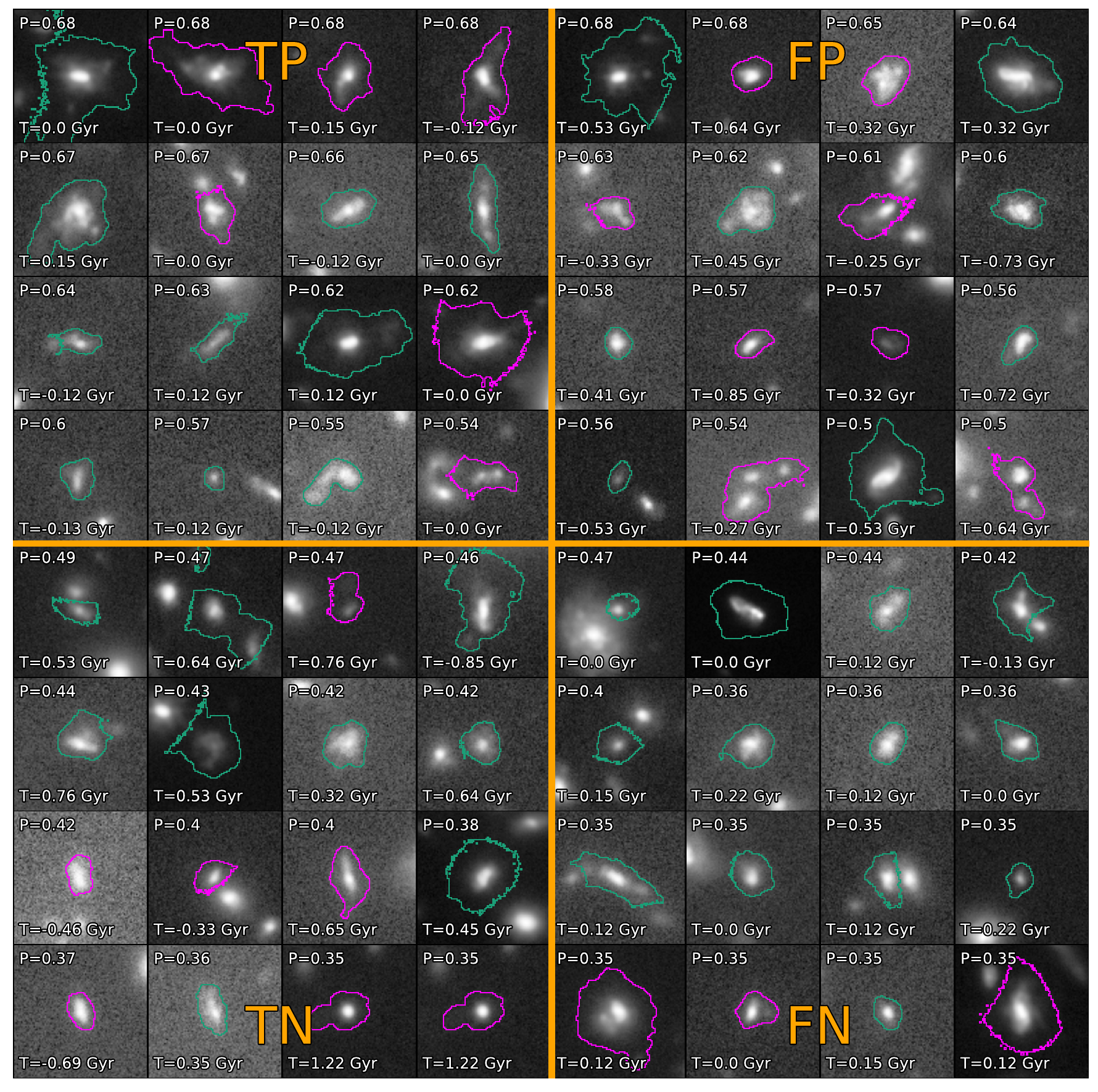}
\caption{Examples of true positive (TP), false positive (FP), true negative (TN), and false negative (FP) (as defined in \S \ref{sec:cnn_sim}) simulated CEERS galaxies in the F356W filter from the $3.5 < z < 4$ redshift bin. Each stamp is 3 x 3 arcsec. Within each stamp, the merger probability output by the DeepMerge network is in the upper left and the timescale since or until the most recent merger (major or minor) is in the bottom left. The outlines show the segmentation map, color-coded by major (\textit{magenta}) and minor (\textit{green}) mergers, respectively.
\label{fig:example_grid_simNN}}
\end{figure*}

Figure \ref{fig:example_grid_simNN} shows the categorization of the mock CEERS test set at $3.5 < z < 4$ into true positives, false positives, true negatives, and false negatives as output by DeepMerge. The top left hand corner of each stamp gives the probability of the object being a merger as determined by the DeepMerge network. The stamps are arranged in order of decreasing
probability, so the horizontal orange line effectively represents the 0.5 probability threshold between merger and non-merger classifications. In each redshift bin, the probabilities range from $\sim 0.3 - 0.7$, which is similar to the range of probabilities output by the random forest.

In Figure \ref{fig:example_grid_simNN}, the bottom left hand corner of each stamp shows the merger timescale for each galaxy, which includes both major and minor mergers. A positive timescale indicates the time since a past merger, and a negative timescale indicates the time until a future merger. Galaxies can have both past and future mergers, so whichever timescale is smallest is shown here. True positives and false negatives will have merger timescales $< 0.25$ Gyr, while false positives and true negatives will have merger timescales $> 0.25$ Gyr. Finally, the segmentation map outlines are color-coded by whether the given timescale corresponds to a major or minor merger. 

\begin{figure*}
\centering
\noindent\makebox[\textwidth][c]{%
    \begin{minipage}{\textwidth}
        \includegraphics[scale=0.36]{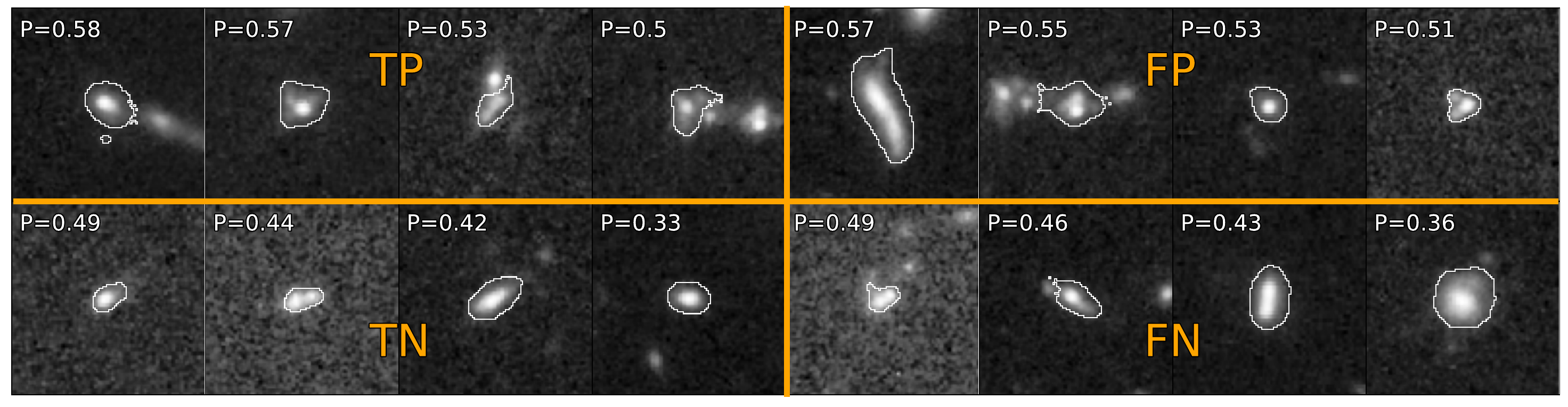}
    \end{minipage}}\hfill
\vskip2ex
\noindent\makebox[\textwidth][c]{%
    \begin{minipage}{\textwidth}
        \includegraphics[scale=0.36]{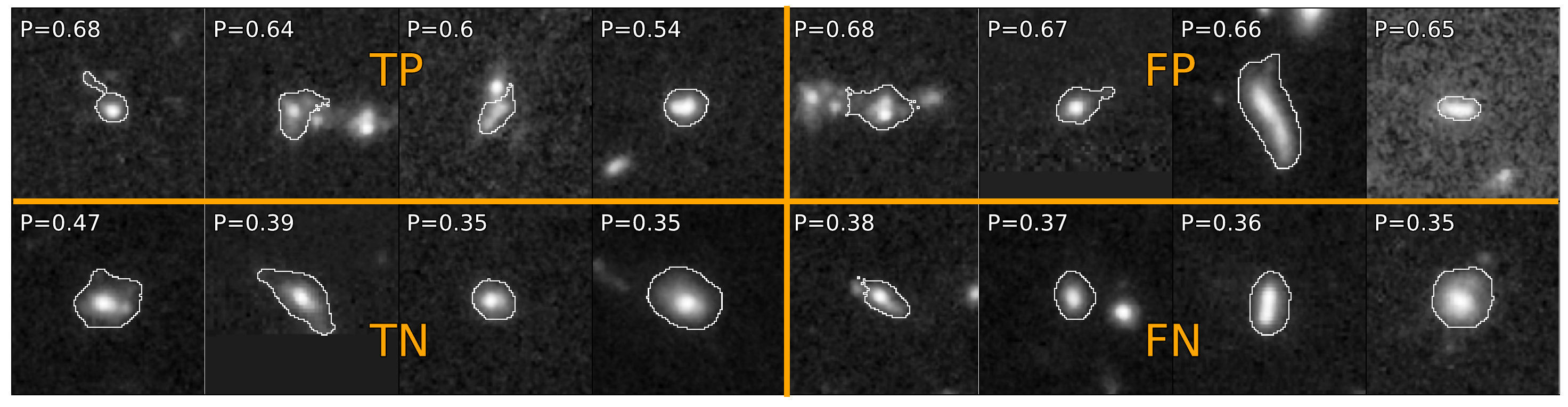}
    \end{minipage}}
\caption{Examples of true positive (TP), false positive (FP), true negative (TN), and false negative (FP) observed CEERS galaxies (from Group 1) in the F356W filter from the $3.5 < z < 4$ redshift bin categorized by the random forest (\textit{\textbf{top}}) or the DeepMerge network (\textit{\textbf{bottom}}). Each stamp is 3 x 3 arcsec. Within each stamp, the merger probability output by the respective algorithm is in the upper left. The white outlines are segmentation map outlines.} \label{fig:grid_obs}
\end{figure*}

Like the false positives in \cite{rose2023}, the false positives in Figure \ref{fig:example_grid_simNN} tend to have irregular segmentation maps due to neighboring or background galaxies, or asymmetric features that are possibly due to a merging event outside of our merger timescale of $\pm 250$ Myr. The average past and future timescales of true negatives are $0.72 \pm 0.30$ Gyr and $0.62 \pm 0.24$ Gyr, respectively. The average past and future timescales of false positives are $0.49 \pm 0.17$ Gyr and $0.41 \pm 0.19$ Gyr, respectively. This indicates that false positives are slightly more likely to be closer in time to a merging event than other non-mergers. On the other hand, the false negatives appear less disturbed, possibly due to minor mergers that would have a minor effect on morphology. The fraction of minor mergers among true positives and false negatives in the simulated CEERS test set is $42.9\%$ and $52.4\%$, respectively, which indicates that false negatives are slightly more likely to be minor mergers. This contrasts with the random forest results in \cite{rose2023} where false positives were not more likely to be closer in time to a merging event and false negatives were not more likely to be minor mergers.

As with the simulated CEERS images, we also categorize the observed CEERS images into true positives, false positives, true negatives, and false negatives. Figure \ref{fig:grid_obs} shows this categorization resulting from the random forest and from the DeepMerge network. In each case, the ``true" label is based on Group 1 visual classification (whether or not two or more people said the object was a merger or interaction). As with Figure \ref{fig:example_grid_simNN}, the probability of the object being a merger as determined by each algorithm is given in the upper left hand corner of each stamp, and stamps are arranged in decreasing probability.

One reason for the poorer performance of these algorithms on the observed data is that the random forest and the DeepMerge network were both trained on data where the ground truth labels were determined by the Illustris past or future merger timescales within $\pm 250$ Myr, while the ground truth labels for the observed CEERS galaxies were based on visual classifications. It is possible that if the algorithms could have been trained on simulated data with labels based on visual classifications, or applied to observed data where the labels were somehow determined by merger timescales, then the algorithms may perform better. Another reason is that the observed galaxies' visual classifications are subjective and were determined by only three classifiers. In some cases, it appears that a different visual classification could be appropriate. For example, the random forest's second false positive in Figure \ref{fig:grid_obs} (or DeepMerge's first false positive in Figure \ref{fig:grid_obs}) could in fact be a merger. Likewise, DeepMerge's second false negative in Figure \ref{fig:grid_obs} could be a non-merger. If these objects had been assigned the opposite visual classification, then the algorithms would have correctly classified them.

\subsection{Merger Rates} \label{sec:merger_rate}

\begin{figure*}[t]
\plotone{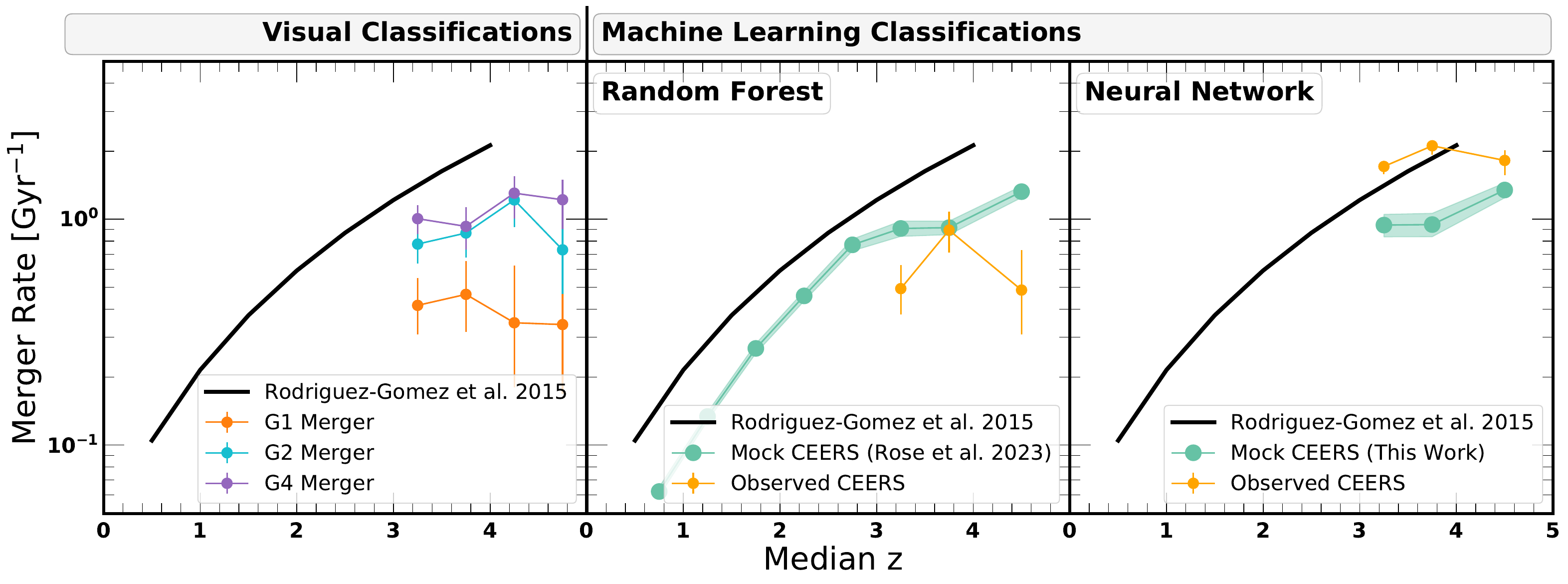}
\caption{Merger rate based on visual classifications (\textbf{\textit{left panel}}), random forest classifications (\textbf{\textit{middle panel}}), and neural network classifications (\textbf{\textit{right panel}}) as compared to
the theoretical Illustris merger rate (\textit{black line}) derived from \cite{rod2015}. The green shaded regions and the observed galaxy error bars
indicate the binomial 95\% confidence interval.
\label{fig:mergerrate}}
\end{figure*}

Figure \ref{fig:mergerrate} shows the simulated CEERS and observed CEERS merger rates. In each panel, the black line is the theoretical Illustris merger rate derived by \cite{rod2015} assuming a merger timescale of 0.5 Gyr, for comparison. The second and third panels show the simulated CEERS merger rate as determined by the random forests (\S \ref{sec:rf_sim}) and by the DeepMerge network (\S \ref{sec:cnn_sim}), respectively, and are in good agreement with each other.

For the DeepMerge-selected simulated CEERS mergers, we calculate the merger rate in each of the three redshift bins. The merger rate was calculated following the same procedure as in \cite{rose2023}, first using the following equation to calculate the merger fraction: 
\begin{equation} \label{eq:mergerrate}
    f_{\mathrm{merger}}(\mathrm{NN}) = \frac{N_{NN}}{N} \frac{PPV}{TPR} <\frac{M}{N}> .
\end{equation} 
Then we divide the merger fraction by our merger timescale of $0.5$ Gyr to obtain the merger rate. Here, $N_{NN}$ is the number of mergers determined by the DeepMerge network and $N$ is the total number of objects in our sample in each redshift bin. PPV is the positive predictive value, or precision, of the classifier. TPR is the true positive rate, or recall, of the classifier. $PPV/TPR$ is meant to correct for the classifier's known incompleteness and purity based on the training set. $<M/N>$ is the average number of merging events per true merger, and accounts for the possibility that mergers can experience more than one merging event within the $\pm 250$ Myr timescale used to determine the labels for the neural network. The DeepMerge simulated CEERS merger rate is shown as the green line in the right panel of Figure \ref{fig:mergerrate}, and is in good agreement with the random forest simulated CEERS merger rate.

The first panel of Figure \ref{fig:mergerrate} shows the observed CEERS merger rate for each visual classification group. This was calculated by first determining the merger fraction using simply:
\begin{equation}
    f_{\mathrm{merger}}(\mathrm{obs}) = \frac{N_{obs}}{N}.
\end{equation} 
We then divide the merger fraction by our timescale of $0.5$ Gyr used for the simulated data. Here, $N_{obs}$ is the number of mergers based on our visual classification groups, and $N$ is the total number of objects in our observed CEERS sample in each redshift bin. The second and third panels also give the observed CEERS merger rate based on the machine learning outputs. These merger rates were calculated as in Equation \ref{eq:mergerrate}, where $PPV/TPR$ is from the respective machine learning algorithm and $<M/N>$ is calculated from the simulated data. Like the simulated merger rates, the observed merger rates in the first and second panel are underestimated compared to the theoretical Illustris merger rate. However, the DeepMerge observed merger rate is higher than the simulated CEERS merger rates, the theoretical Illustris merger rate, and the observed merger rates in the first panel due to the fact that DeepMerge overestimated the number of mergers. The random forest observed merger rate is in better agreement with the observed merger rates in the first panel. In addition, the observed merger rates appear relatively flat while the simulated mergers rates increase at $z > 4$.

\subsection{Comparison with Previous Works} \label{sec:lit_comp}
 
As discussed in \S \ref{sec:ch3intro}, few studies apply convolutional neural networks to the task of merger identification at high redshift. We compare our work to \cite{cip2020}, \cite{fer2020}, and \cite{pearson2019_gm20}. The DeepMerge network is introduced in \cite{cip2020} to identify $z=2$ simulated Illustris-1 HST mergers. They have access to four different ``camera'' perspectives, which they use to create a sample of 8930 unique galaxies. They then augment the merger class to create a more balanced sample of 8120 mergers and 7306 non-merger before they split their data into train and test sets. Their merger class consists of mergers with a stellar mass ratio of at least 0.1 and within a time frame of $\pm 250$ Myr. They find that DeepMerge achieves an accuracy of $76\%$ when training and testing on noisy data. They report correctly classifying $78\%$ of mergers and $73\%$ of non-mergers in their noisy test set. In \cite{cip2021}, they use domain adaption techniques to apply their trained network to real data; however, this is done at $0.005 < z < 0.1$ using data from the Galaxy Zoo project \citep{lin2008,lin2011}.

\cite{fer2020} use simulated IllustrisTNG300-1 CANDELS major mergers (up to $z\sim3$) within $\pm 0.3$ Gyr and non-mergers. Additionally, they impose a limit on the separation between each pair of galaxies. With these criteria, they curate a sample of $\sim30,000$ distinct major merger candidates. For the non-mergers, they impose a timescale limit of $\pm 0.5$ Gyr (i.e., the galaxy must have a merger timescale \textit{outside} this time frame). Using these criteria, they build up a random sample of non-mergers that equals the number of mergers in each redshift bin. They train two classifiers -- one to identify mergers vs. non-mergers, and a second to further subdivide mergers into pre- and post-mergers -- and are able to correctly classify $87\%$ and $94\%$ or pre- and post-mergers, respectively. They apply their methodology to a sample of 3759 real CANDELS galaxies with visual classifications given by \cite{kart2015}, but do not report an accuracy for this test. They do note that since their validation sample is balanced while the CANDELS sample is not, the performance of the model on simulated data will not translate to the same performance of the model on real data.

Finally, \cite{pearson2019_gm20} both train and test their convolutional neural network on observed CANDELS galaxies up to $z < 4$, in the F814W, F125W, and F160W bands, using visual classifications from \cite{kart2015}. They construct training samples of 694 objects in each class in each of their five redshift bins from $0 < z < 4$. They also augment their images to create a larger training set. They report a final accuracy of $81.8\%$ after testing on the full redshift range of $0 < z < 4$.

Each of these studies reports better performance than we report in this work, which could be due to a number of differences. First, \cite{cip2020} and \cite{fer2020} use Illustris-1 and IllustrisTNG300-1, respectively, while we use IllustrisTNG100-1. \cite{fer2020} impose additional criteria to define their merger sample and non-merger samples. Perhaps most importantly, both \cite{cip2020} and \cite{fer2020} are able to curate a large training set, even before augmentation. All three studies also use balanced training, validation, and test sets, whereas our validation and test sets are left unbalanced since real data will be unbalanced. Additionally, \cite{cip2020} augment their data prior to splitting their data into training, validation, and test sets, and it is unclear if they ensured that all augmentations of a given galaxy end up in the same subset. Otherwise, the training, validation, and test sets would be correlated which could inflate the accuracy of the network. \cite{fer2020} do not specifically comment on the performance of their first classifier, which separates mergers from non-mergers. \cite{pearson2019_gm20} do not specifically report performance on their non-merger class. Thus these papers may have focused on optimizing performance to correctly identify most mergers, rather than focusing on the purity of the merger and non-merger classifications, similarly to \cite{sny2019}. Finally, \cite{cip2020} and \cite{fer2020} study galaxies at $z < 3$ with HST images while we push to $3 < z < 5$ with JWST. \cite{pearson2019_gm20} do extend out to $z = 4$, but their training set consists for data from the full range of $0 < z < 4$.

\section{Conclusions} \label{sec:conclude}

In this work, we investigate using the DeepMerge convolutional neural network algorithm to identify merging galaxies at $3 < z < 5$ in simulated CEERS images constructed from IllustrisTNG and the Santa Cruz SAM. Additionally, we calculate morphology parameters for observed CEERS galaxies using the morphology programs \texttt{Galapagos-2} and \texttt{statmorph}. We apply the trained random forests from \cite{rose2023}, a $4 < z 
 < 5$ random forest trained in this work, and the DeepMerge neural network trained in this work to observed CEERS galaxies, and verify using visual classifications from \cite{kart2023}. We find the following results:
\begin{enumerate}
\item The $4 < z < 5$ random forest performed similarly to those in \cite{rose2023}, correctly classifying $59\%$ of non-merging galaxies and $67\%$ of merging galaxies.
\item The DeepMerge network correctly classified $\sim60 - 70\%$ of simulated merging and non-merging galaxies. Although Grad-CAMs from our tests using six filters were not clearly interpretable, Grad-CAMs from our tests using one filter showed that the networks usually activated around the edges of galaxies as expected.
\item The random forests and neural networks did not perform as well when applied to real CEERS galaxies. In particular, the neural network tended to classify most objects ($60 - 80\%$) as mergers. This could be due to a number of reasons, such as the different methods of defining mergers for the simulated data and the observed data.
\item As in \cite{rose2023}, DeepMerge false positives tend to appear to have close neighbors or asymmetric features, and false negatives appear less disturbed. The false positives appear to be more likely to be closer in time to merging events than the true negatives and false negatives appear more likely to be minor mergers than true positives. The categorization of observed CEERS galaxies shows that positive classifications do look more irregular and asymmetric and negative classifications do look more undisturbed, but also that visual classifications are subjective.
\item Merger rates from (a) observed visual classifications only, (b) RF-selected observed merger classifications, and (c) DeepMerge-selected simulated merger classifications generally agree with each other and with the RF-selected simulated merger rate from \cite{rose2023}. The merger rate calculated using DeepMerge-selected observed merger classifications is overestimated comparatively, due to the fact that DeepMerge misclassified almost all observed non-mergers.
\end{enumerate}

\subsection{Future Work}

One clear area of improvement is curating a larger sample of simulated mergers in order to increase the number of objects in the training set which could help the network better learn merger signatures. This could be done by remaking the pristine simulated images using the larger IllustrisTNG300-1 simulation \citep[as in][]{fer2020} or using another large volume, high resolution simulation. A larger sample would also enable the performance as a function of other physical parameters that can impact the observed merger signatures, such as the gas fraction of the galaxies, their merger ratio, and their relative orbits. Another area of improvement would be investigating the use of transfer learning techniques to improve performance on the observed simulated CEERS data. \cite{cip2021} show that using transfer learning techniques improves the performance of their simulation-trained networks on observed data by up to $\sim 20\%$. Finally, it would be interesting to train and test machine learning techniques directly on observed CEERS galaxies \citep[similar to][]{pearson2019_gm20} once more visual classifications become available. It would also be interesting to compare the Grad-CAM maps to the regions of the images for which the morphology parameters are sensitive, in order to better understand the features that the neural network has learned.

\begin{acknowledgments}
Support for this work was provided by NASA through grants JWST-ERS-01345.015-A and HST-AR-15802.001-A awarded by the Space Telescope Science Institute, which is operated by the Association of Universities for Research in Astronomy, Inc., under NASA contract NAS 5-26555. This research is based in part on observations made with the NASA/ESA Hubble Space Telescope obtained from the Space Telescope Science Institute, which is operated by the Association of Universities for Research in Astronomy, Inc., under NASA contract NAS 5–26555. This work is based in part on observations made with the NASA/ESA/CSA James Webb Space Telescope. The data were obtained from the Mikulski Archive for Space Telescopes at the Space Telescope Science Institute, which is operated by the Association of Universities for Research in Astronomy, Inc., under NASA contract NAS 5-03127 for JWST. These observations are associated with program \#1345.

The authors acknowledge Research Computing at the Rochester Institute of Technology for providing computational resources and support that have contributed to the research results reported in this publication \citep{ritcomp}

The authors acknowledge the Texas Advanced Computing Center (TACC) at The University of Texas at Austin for providing HPC resources that have contributed to the research results reported within this paper (\href{http://www.tacc.utexas.edu}{http://www.tacc.utexas.edu}).

\end{acknowledgments}

\vspace{5mm}
\facilities{JWST (NIRCam)}

\software{Source Extractor \citep{ber1996}, Galapagos-2 \citep{hau2013}, statmorph \citep{rod2019}, DeepMerge \citep{cip2020}}

\appendix

\section{Confusion matrices for observed CEERS galaxies}
\begin{figure}
\minipage{\textwidth}%

  \includegraphics[width=0.33\linewidth]{conmat_3p0to3p5_group1.pdf}
  \includegraphics[width=0.33\linewidth]{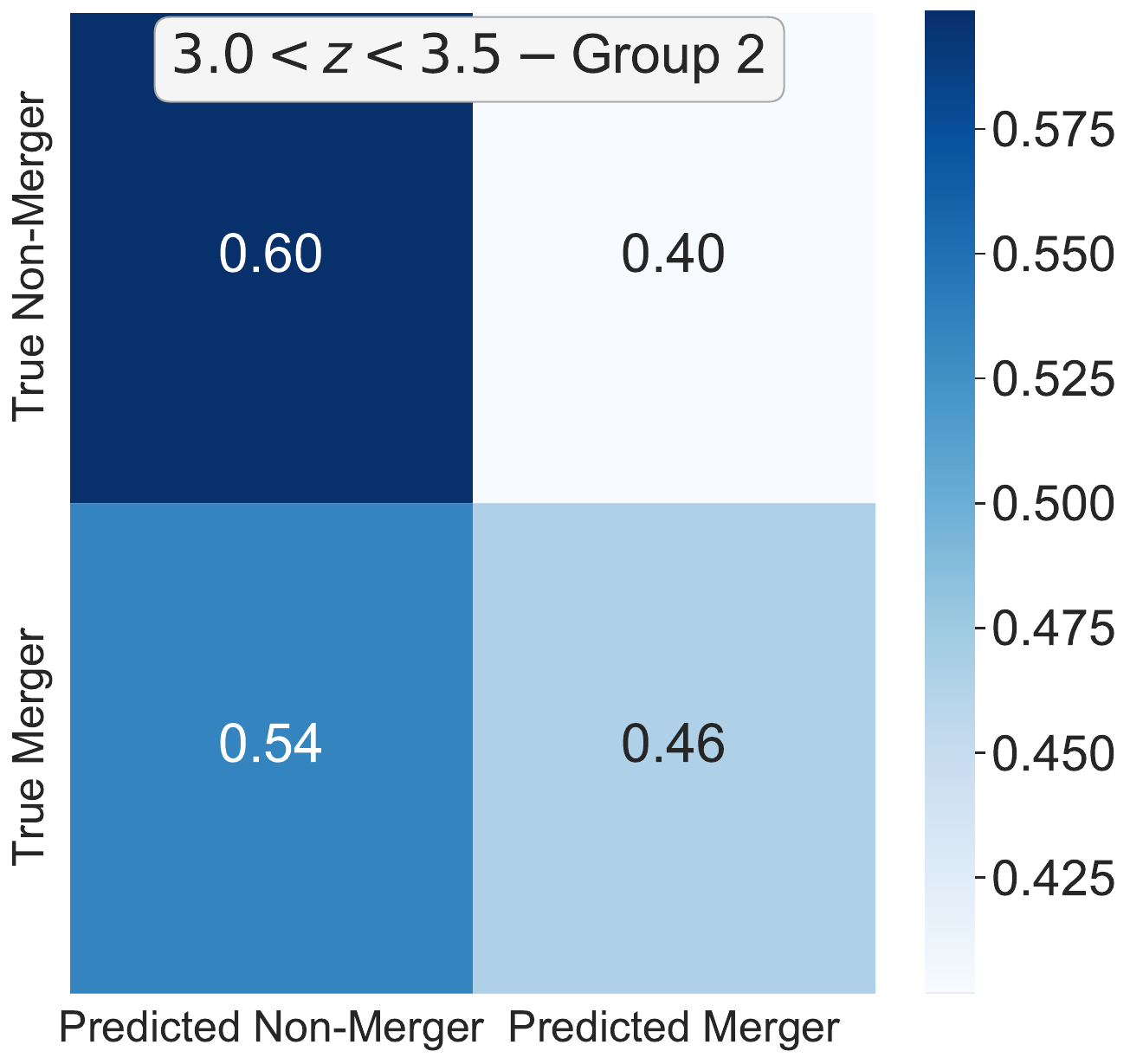}
  \includegraphics[width=0.33\linewidth]{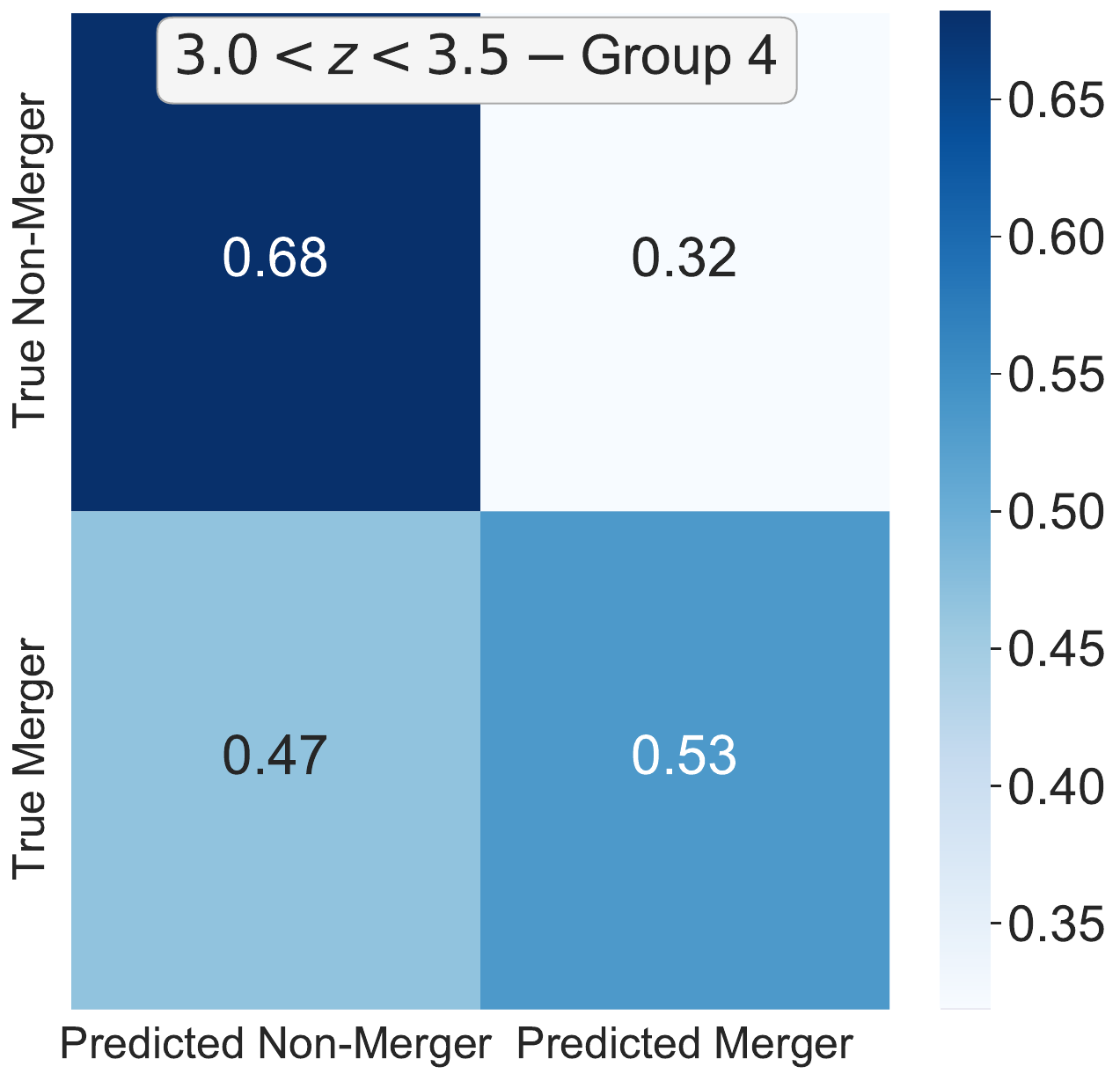}
\endminipage\hfill
\vskip2ex
\minipage{\textwidth}%
  
  \includegraphics[width=0.33\linewidth]{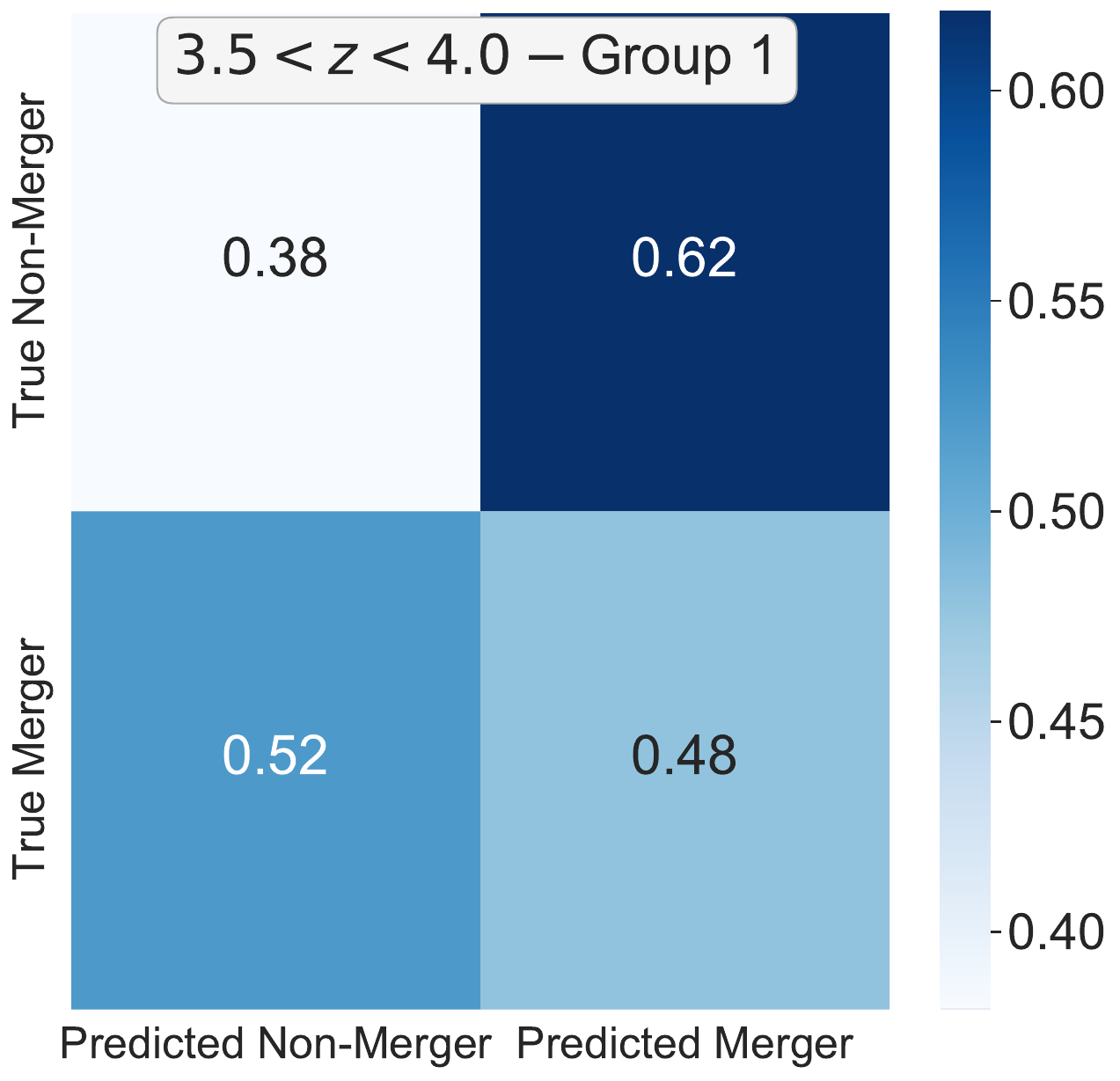}
  \includegraphics[width=0.33\linewidth]{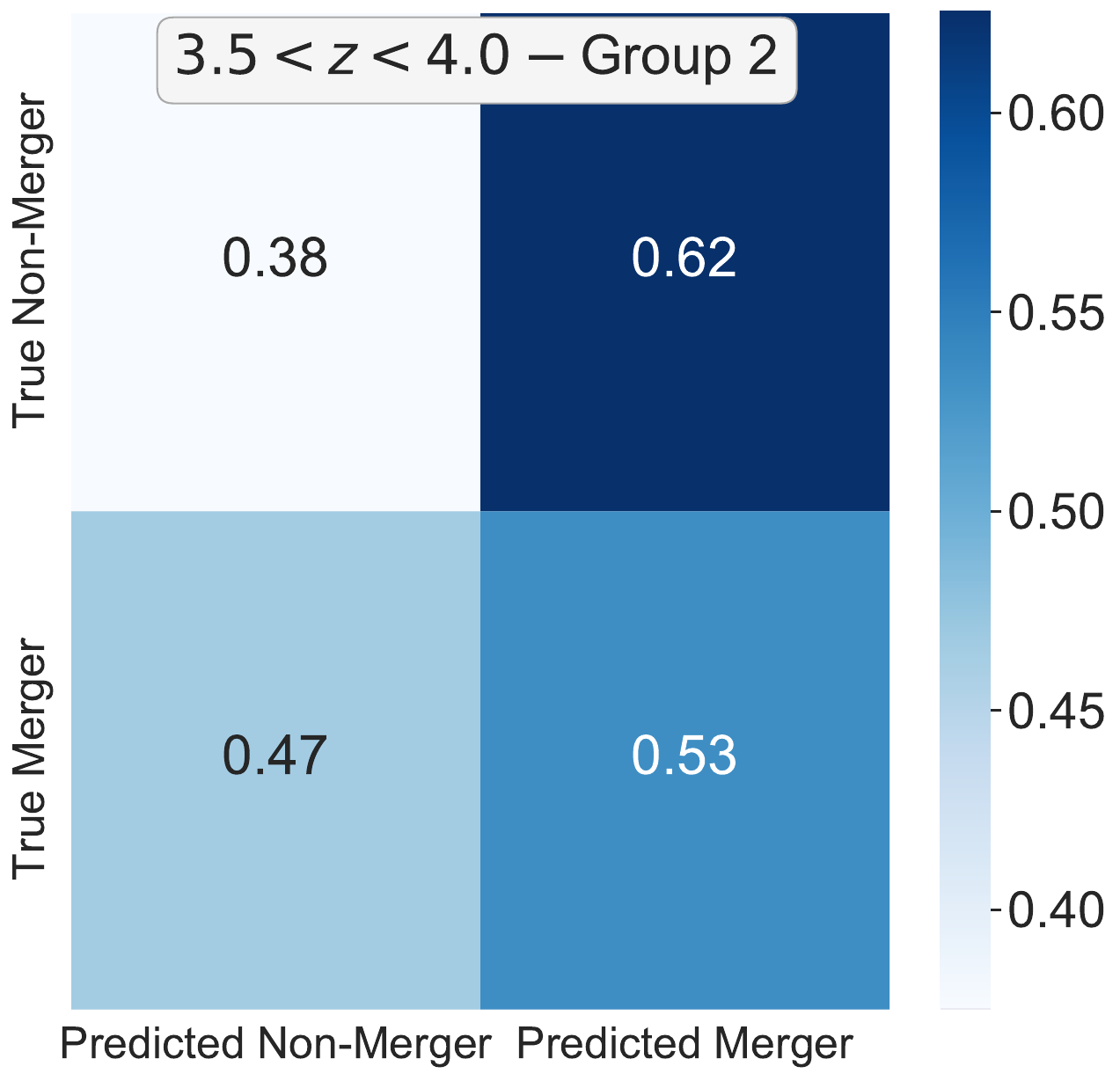}
  \includegraphics[width=0.33\linewidth]{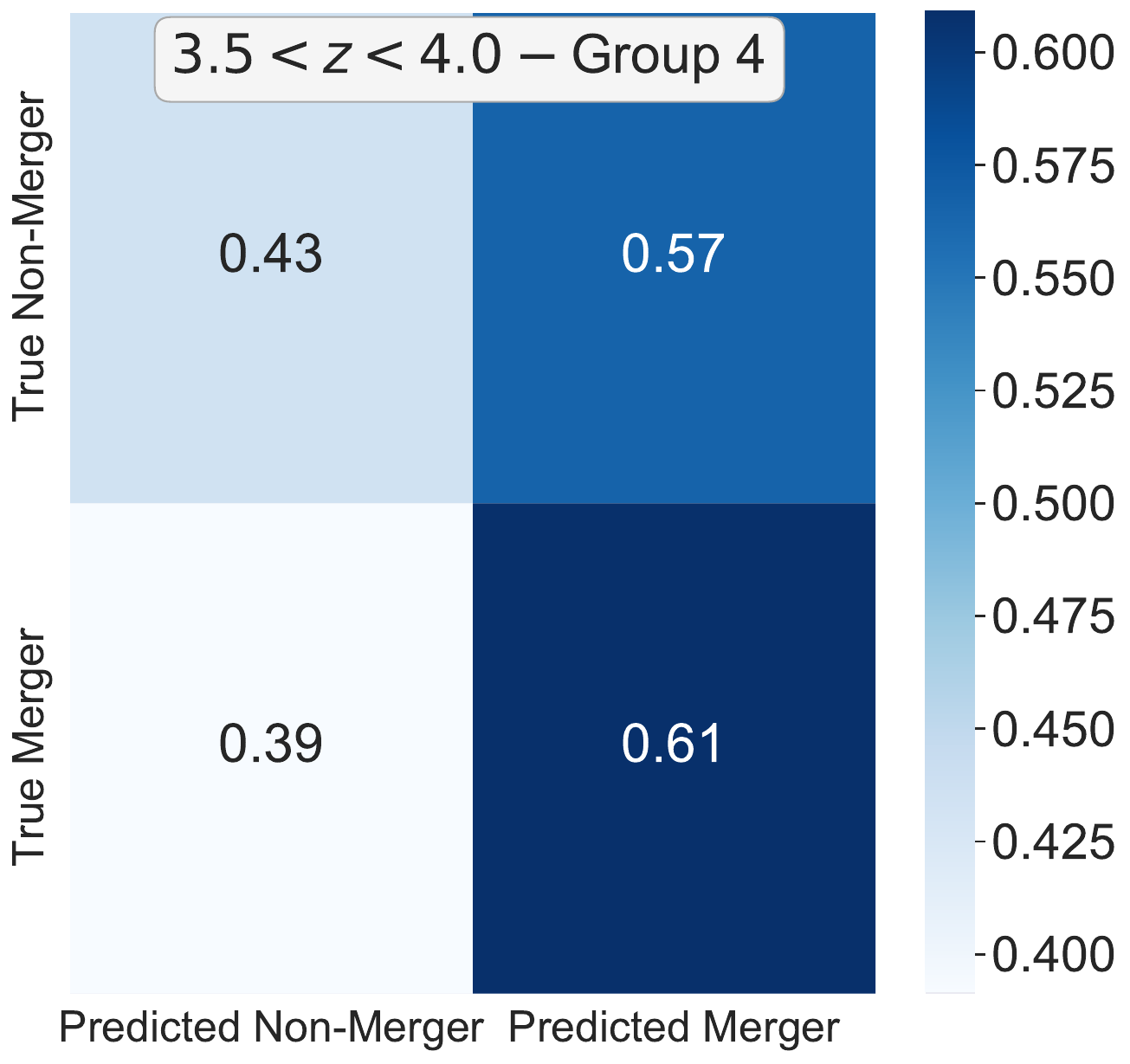}
  
\endminipage\hfill
\vskip2ex
\minipage{\textwidth}%
  
  \includegraphics[width=0.33\linewidth]{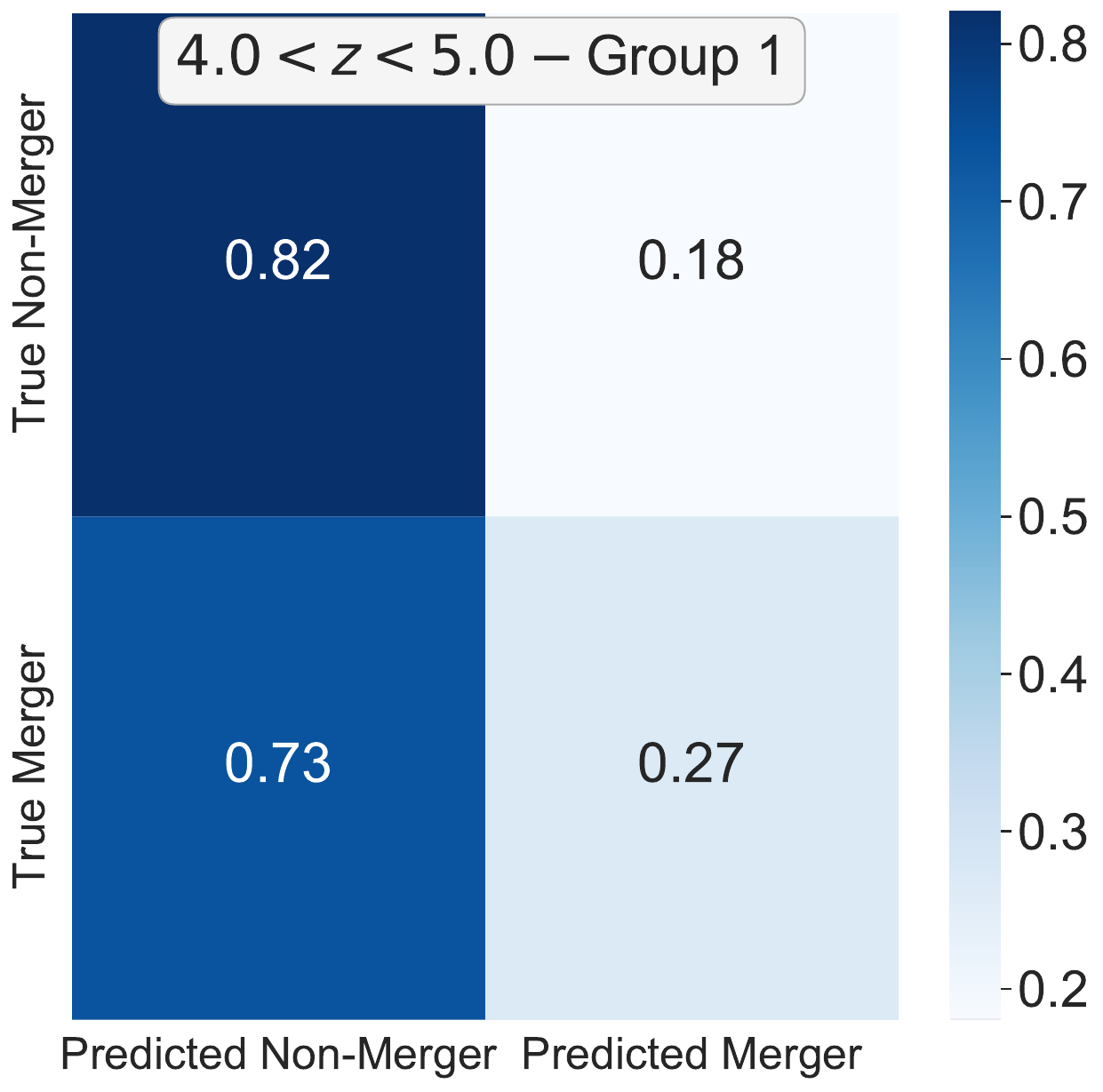}
  \includegraphics[width=0.33\linewidth]{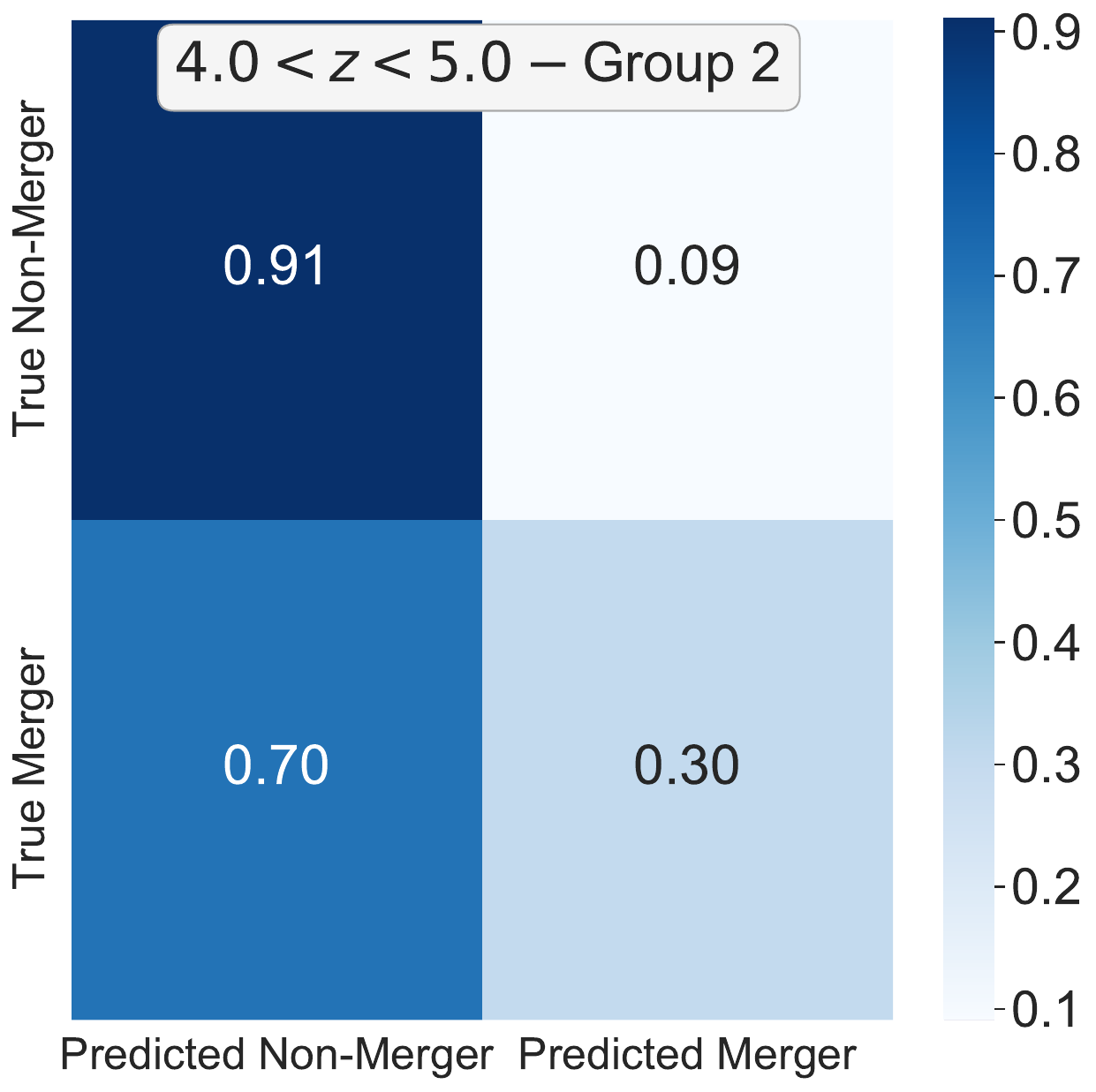}
  \includegraphics[width=0.33\linewidth]{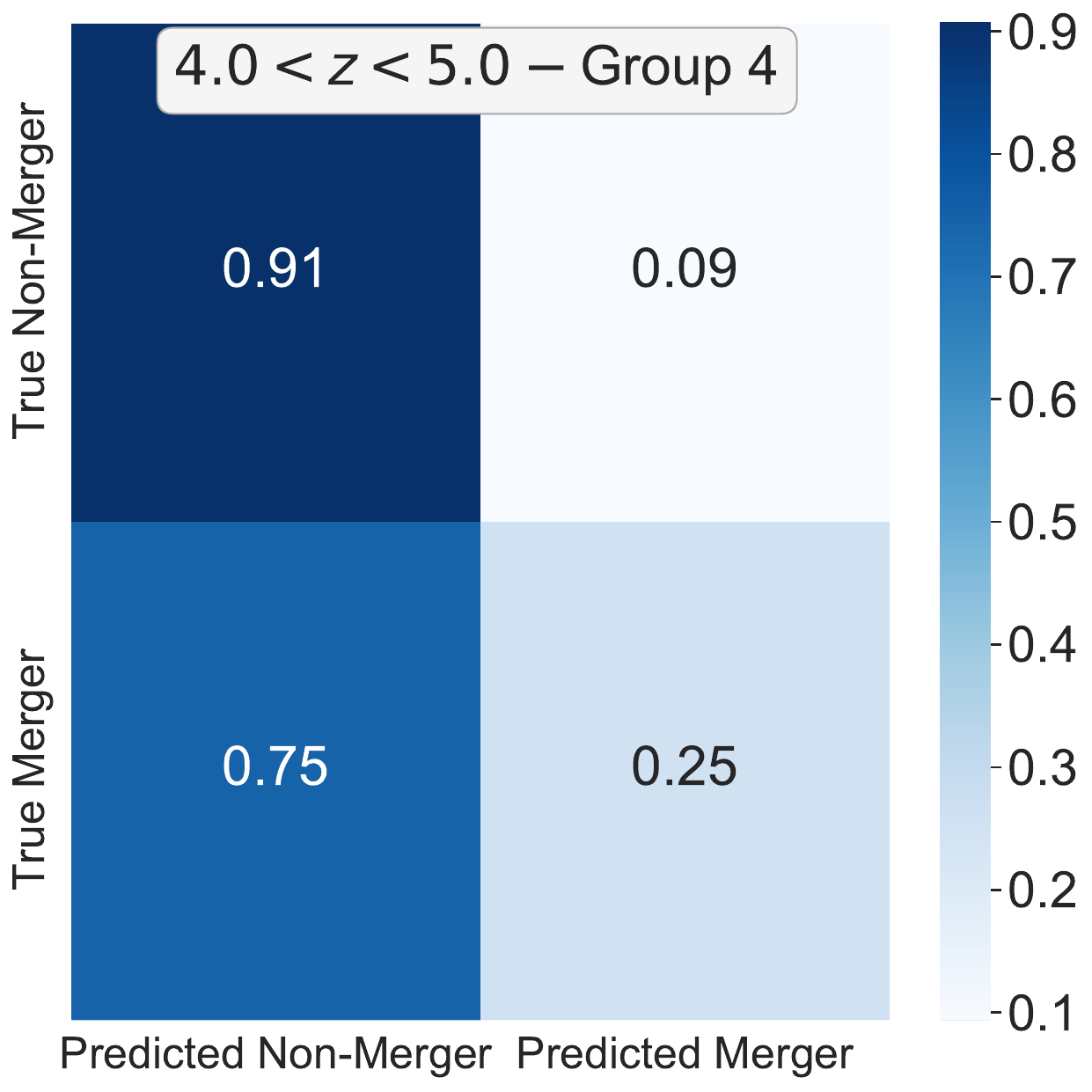}
  
\endminipage
\caption{Random forest confusion matrices for observed CEERS galaxies at $3 < z < 3.5$ (\textbf{\textit{top row}}), at $3.5 < z < 4$ (\textbf{\textit{middle row}}), and at $4 < z 
<5$ (\textbf{\textit{bottom row}}).} \label{fig:rf_real_cm}
\end{figure}

\begin{figure}
\minipage{\textwidth}%
  
  \includegraphics[width=0.33\linewidth]{conmat_NN_3p0to3p5_group1_isabella.pdf}
  \includegraphics[width=0.33\linewidth]{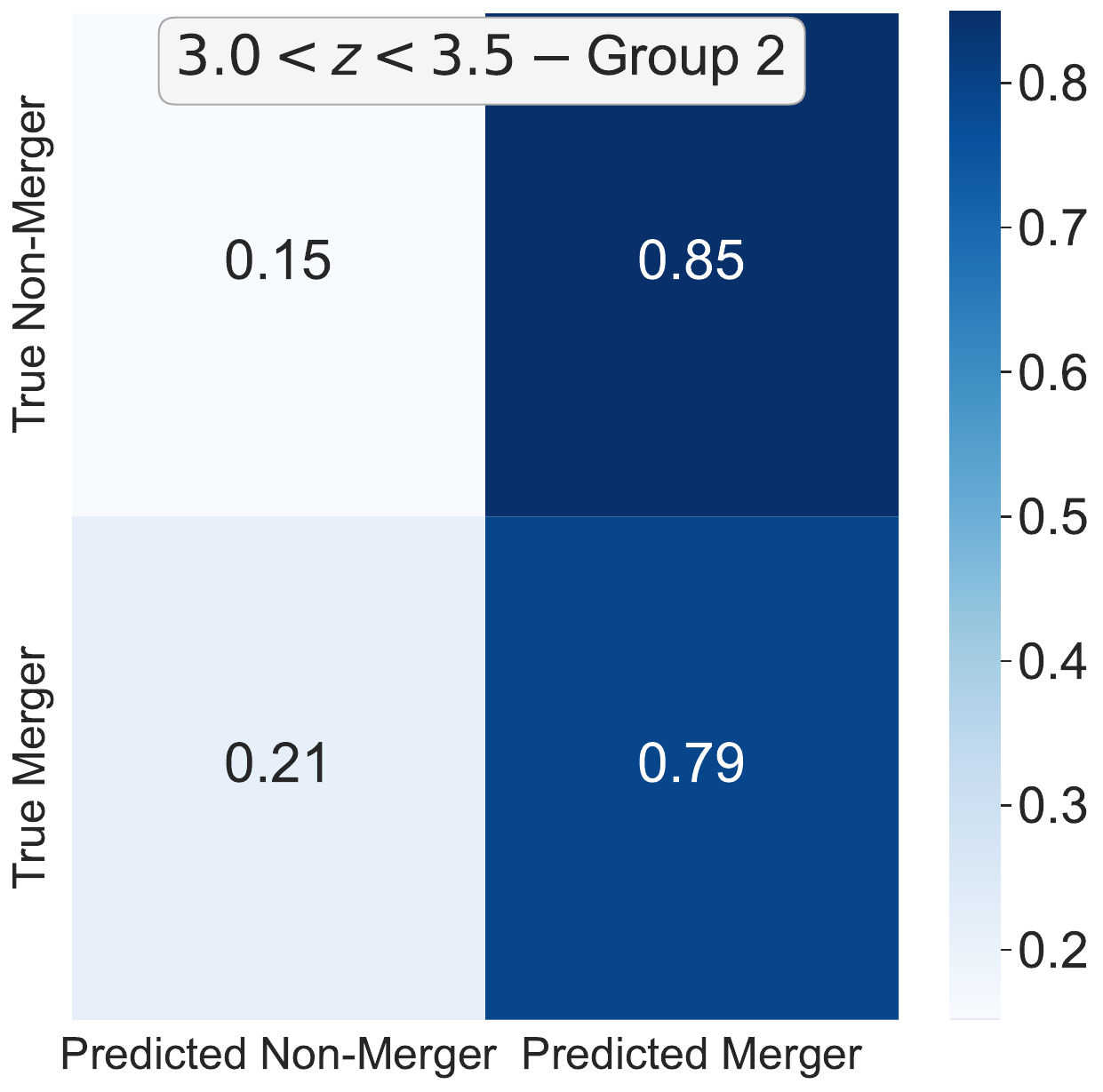}
  \includegraphics[width=0.33\linewidth]{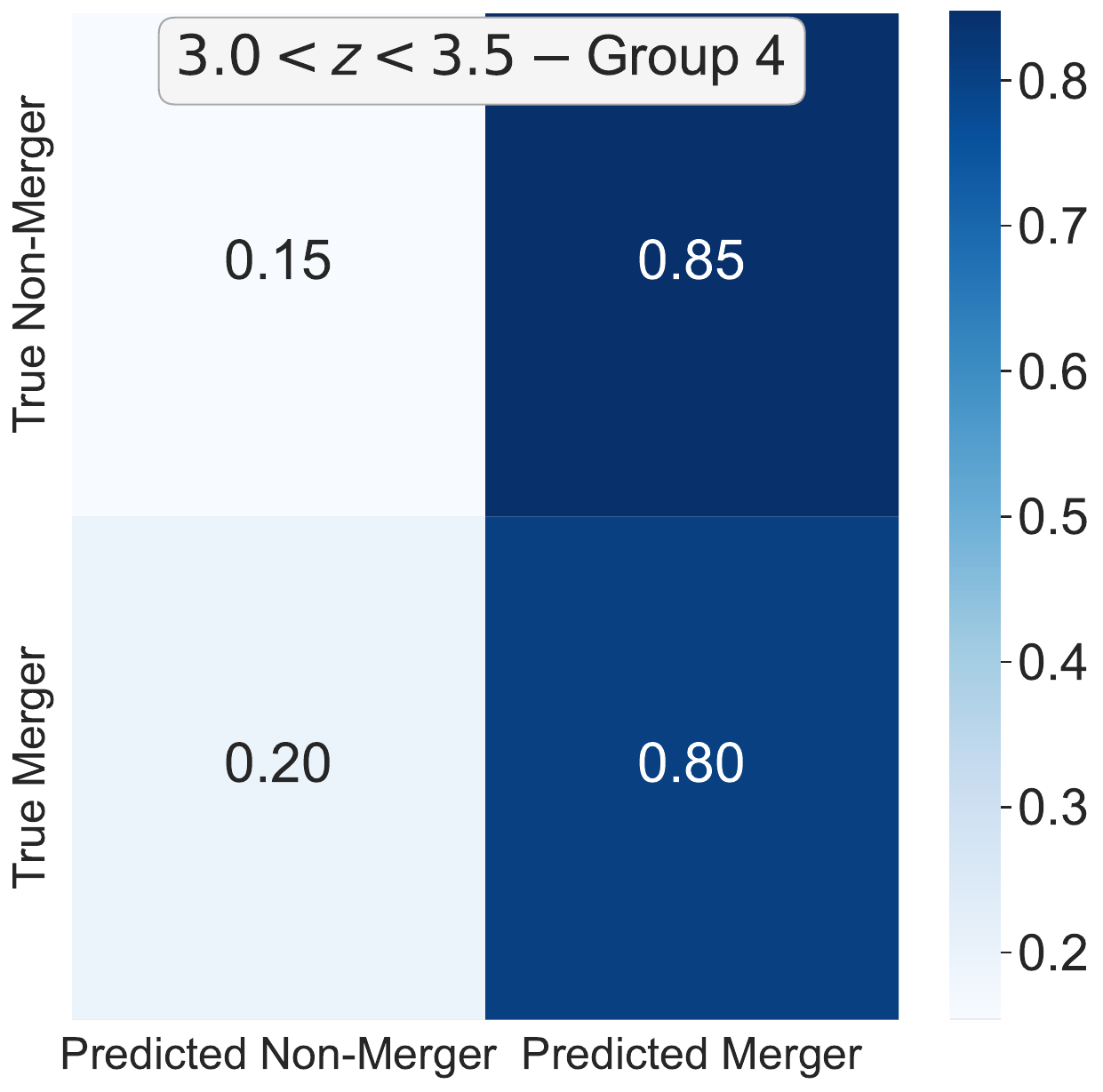}
\endminipage\hfill
\vskip2ex
\minipage{\textwidth}%
  
  \includegraphics[width=0.33\linewidth]{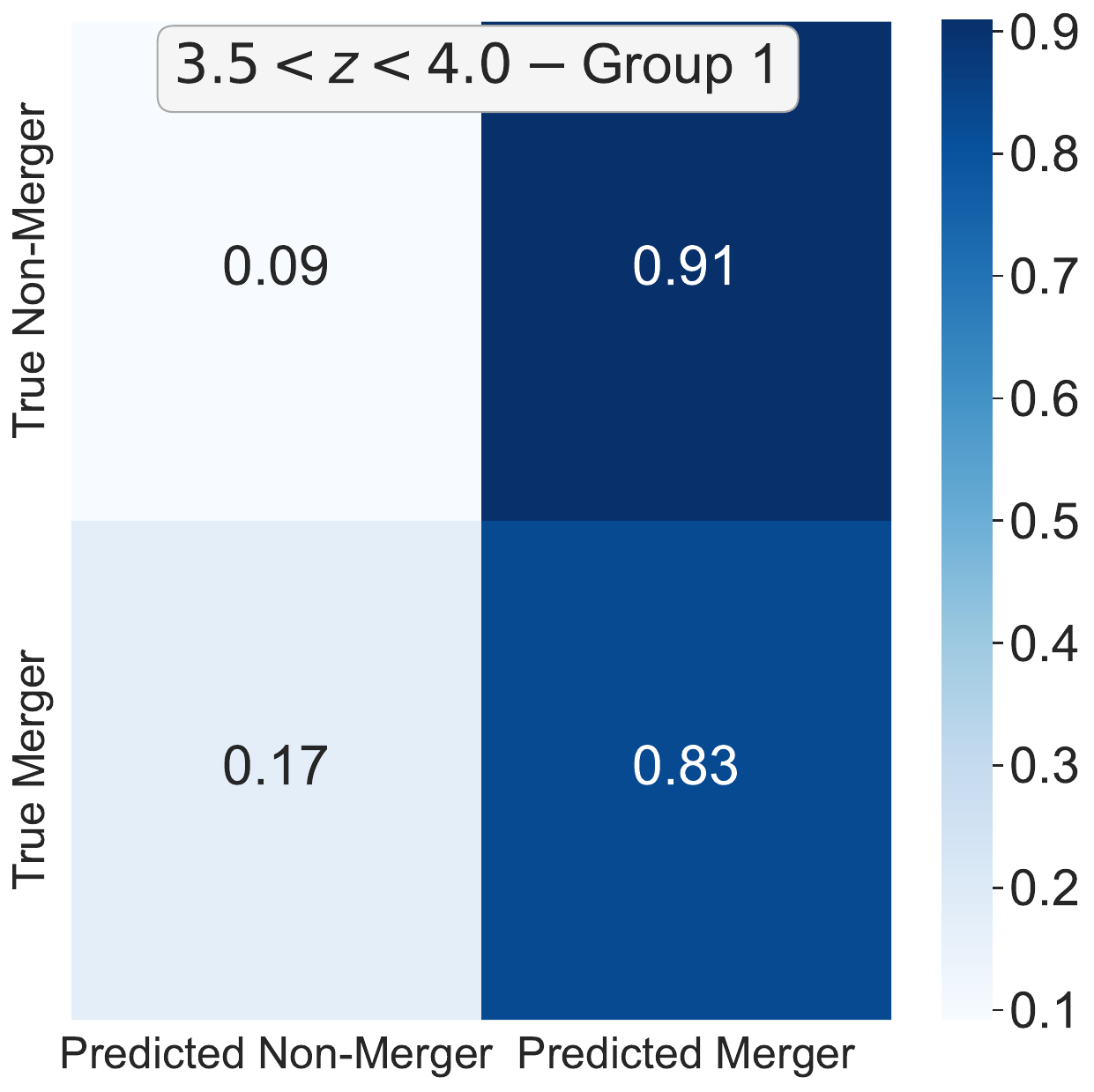}
  \includegraphics[width=0.33\linewidth]{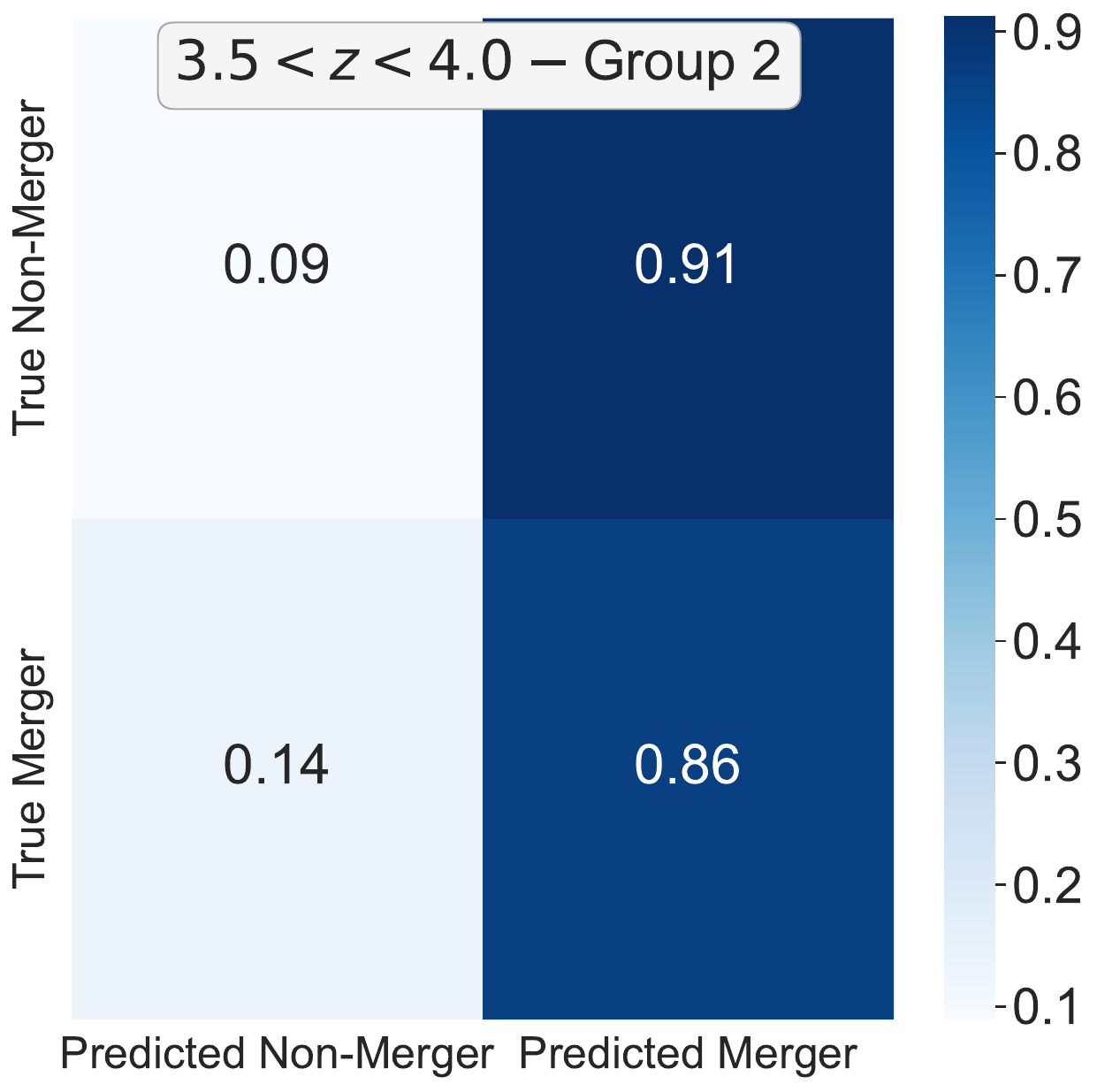}
  \includegraphics[width=0.33\linewidth]{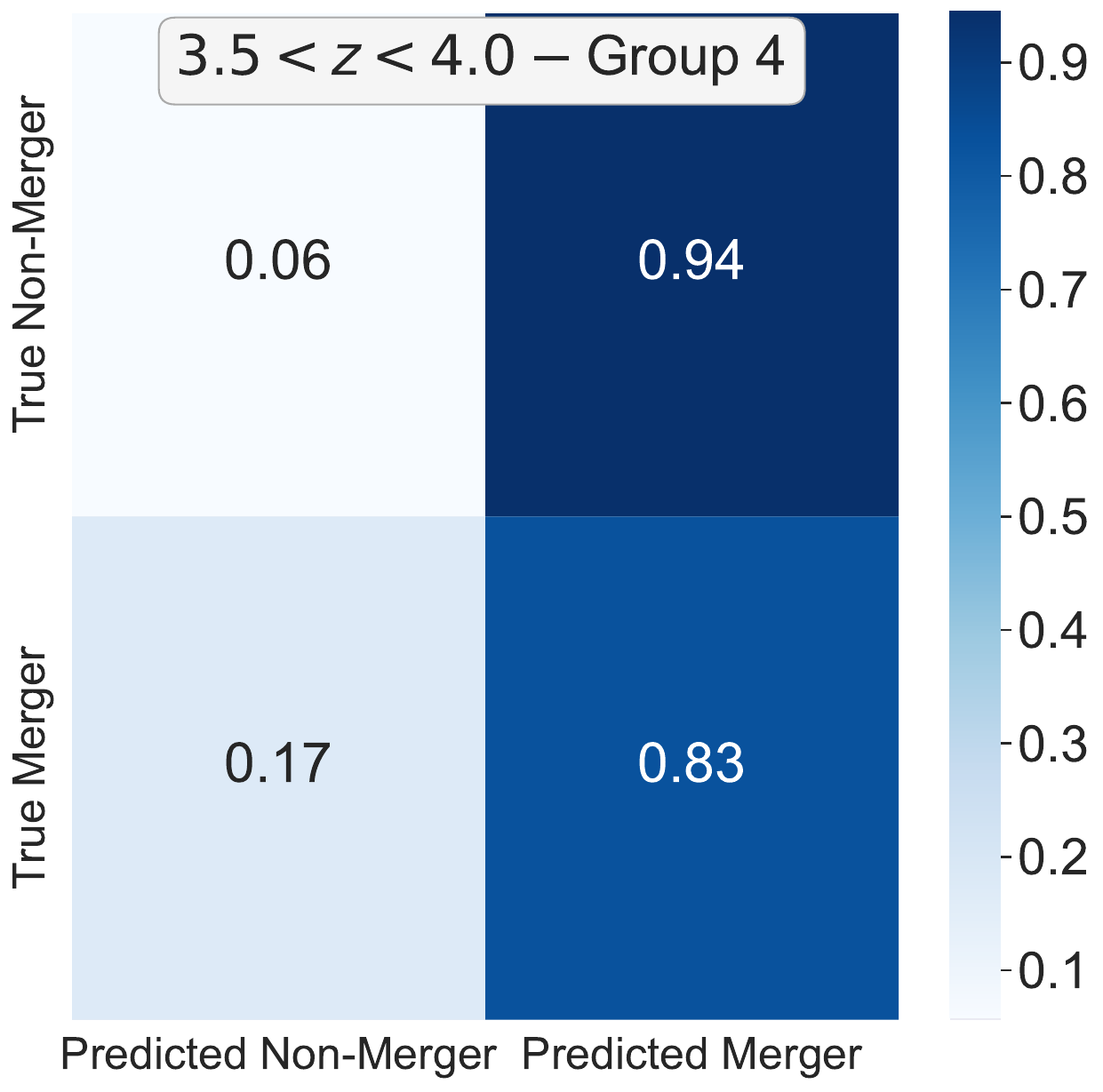}
  
\endminipage\hfill
\vskip2ex
\minipage{\textwidth}%
  
  \includegraphics[width=0.33\linewidth]{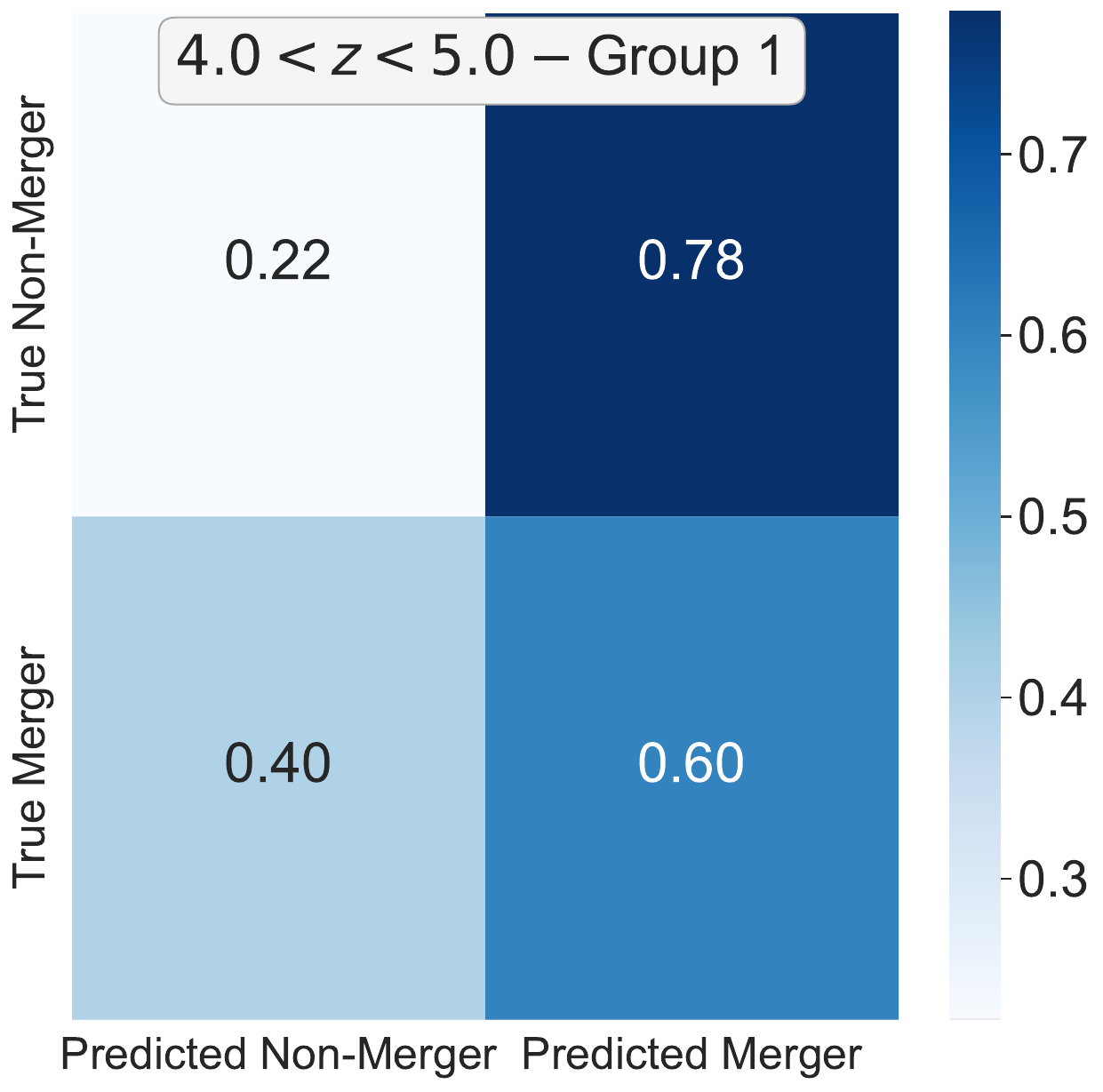}
  \includegraphics[width=0.33\linewidth]{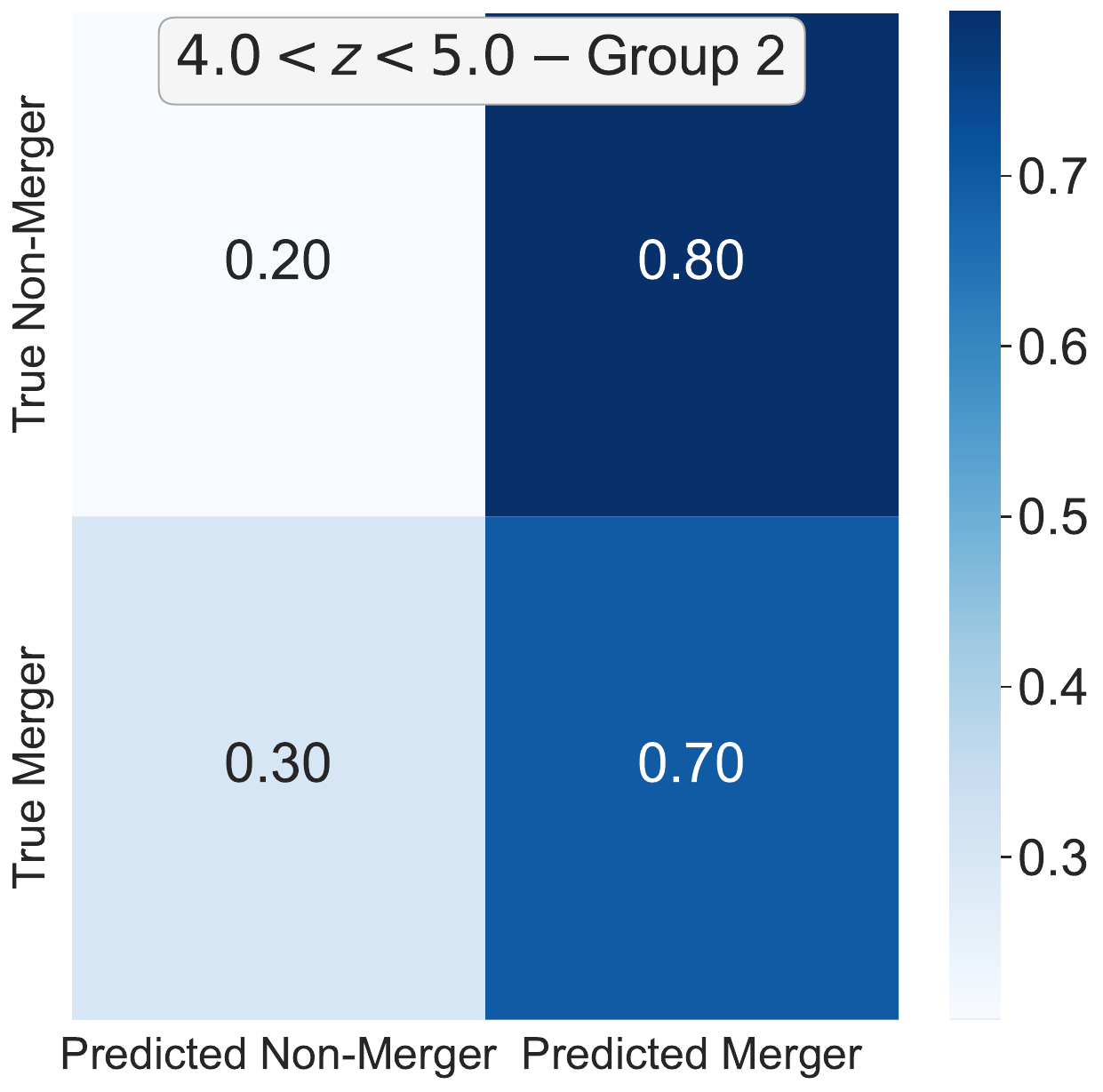}
  \includegraphics[width=0.33\linewidth]{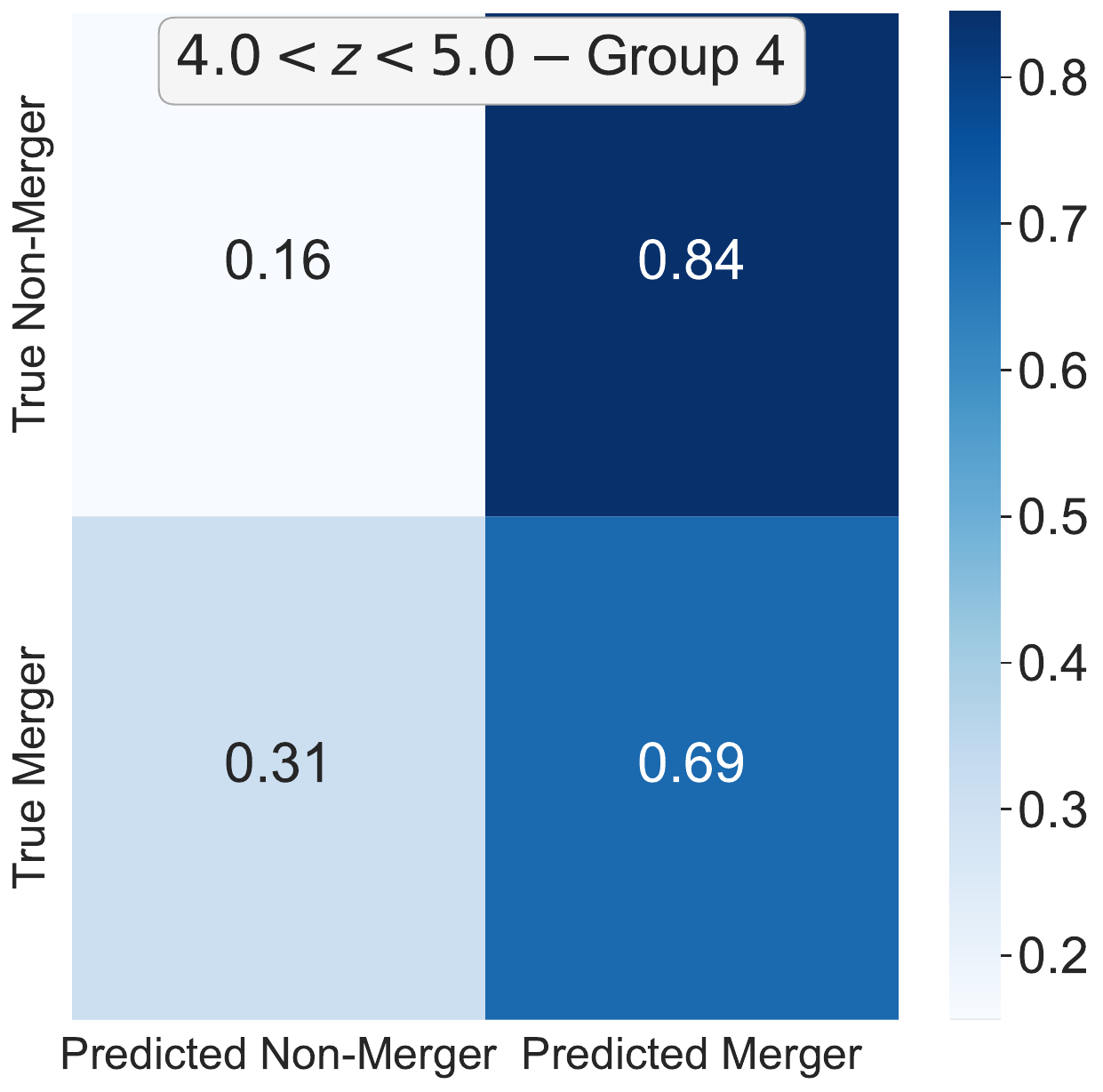}
  
\endminipage
\caption{DeepMerge neural network confusion matrices for observed CEERS galaxies at $3 < z < 3.5$ (\textbf{\textit{top row}}), at $3.5 < z < 4$ (\textbf{\textit{middle row}}), and at $4 < z < 5$ (\textbf{\textit{bottom row}}).} \label{fig:nn_real_cm}
\end{figure}

Figure \ref{fig:rf_real_cm} shows the results of applying the trained random forests to observed CEERS galaxies in \S \ref{sec:rf_real}. Figure \ref{fig:nn_real_cm} shows the results of applying the trained DeepMerge neural networks to observed CEERS galaxies in \S \ref{sec:cnn_real}.

\bibliography{rose_references.bib}{}
\bibliographystyle{aasjournal}

\end{document}